\documentstyle[epsf]{article}
\def\ket#1{\vert#1\rangle}
\def\bra#1{\langle#1\vert}
\def\brak#1#2{\langle#1\vert #2\rangle}

\def\bx{{\bf{x}}}
\def\bp{{\bf{p}}}
\def\bk{{\bf{k}}}
\def\bpi{{\bf{\pi}}}
\def\bxi{{\bf{\xi}}}
\def\bq{{\bf{q}}}

\def\axd{\hat{ a}^{\dag} (\bx)}
\def\apd{\hat a^{\dag} (\bp)}
\def\ax{\hat{ a}^{} (\bx)}
\def\ap{\hat  a^{} (\bp)}

\def\be{\begin{equation}}
\def\ee{\end{equation}}
\def\bea{\begin{eqnarray}}
\def\eea{\end{eqnarray}}
\def\ri{\right)}
\def\lef{\left(}
\def\dst{\displaystyle\phantom{|}}
\def\ov{\over\dst}
\def\om{\omega}

\title{ Multi-Boson Correlations Using Wave-Packets
}
\author{
J. Zim\'anyi$^1$ and T. Cs\"org\H o$^{1,2}$ \\
{\it $^1$MTA KFKI RMKI, H-1525 Budapest 114. POB. 49, Hungary}\\
{\it $^2$Department of Physics, Columbia University,\\
 538 W 120-th Street, New York, NY 10027 }
 }
\date{May 13, 1997}
\begin{document}
\rightline{CU-TP-818R/1997}
\bigskip
\bigskip
\begin{center}
{\Large
 Multi-Boson Correlations Using Wave-Packets
}
\end{center}
\medskip
\begin{center}
{\sc J. Zim\'anyi$^1$\thanks{\rm
	E-mails: jzimanyi@sunserv.kfki.hu, csorgo@sunserv.kfki.hu}
	and T. Cs\"org\H o$^{1,2}$}
\end{center}

\begin{center}
{\it $^1$MTA KFKI RMKI, H-1525 Budapest 114. POB. 49, Hungary}\\
{\it $^2$Department of Physics, Columbia University,\\
 538 W 120-th Street, New York, NY 10027 }
\end{center}
\bigskip\bigskip
\date
\bigskip\bigskip
\begin{abstract}
  Brooding over bosons, wave packets and Bose - Einstein
correlations, we present a generic quantum mechanical
system that contains arbitrary number of bosons
characterized by wave-packets and that can undergo
a Bose-Einstein condensation either by cooling, or increasing the number
density of bosons, or by increasing the overlap of the
multi-boson wave-packet states, achieved by changing the
size of the single-particle wave-packets.
We show that the $n$-particle correlations may mimic coherent or
chaotic behavior for certain limiting wave-packet sizes.
Effects of complete $n$-particle symmetrization are included.
The resulting weights which fluctuate between 1 and $n!$ are summed
up with the help of a formal analogy between the considered wave-packet
system and an already explored multi-boson plane-wave system.
We solve the model analytically in the highly condensed
and in  the rare gas limiting cases, numerically
in the intermediate cases. The relevance of the model
to  multi-pion production in high energy heavy ion physics
as well as to the Bose-Einstein condensation of atomic vapors
is discussed. As a by-product, a new class of probability distribution
functions is obtained and the critical density for the onset of
pion-lasing is derived. The multiplicity dependence of single-particle
momentum distributions are predicted that could be utilized in 
future event-by-event measurements.

\end{abstract}
\bigskip\bigskip
\section{Introduction \label{s:1}}
The study of the statistical properties of quantum systems has a long
history with important recent developments. In high energy physics,
quantum statistical correlations are studied in order to infer the
space-time dimensions of the elementary particle reactions.
The measured characteristic length scales are on the $10^{-15}$ m scale,
the time scales are of the order of  $10^{-23}$ sec. In high energy heavy
ion collisions hundreds of bosons are created in the present CERN SPS
reactions when $Pb + Pb$ reactions are measured at 160 AGeV laboratory
bombarding energy. At the planned RHIC accelerator, thousands of pions
could be produced in a unit rapidity interval. If the number of bosons
in a unit value of phase-space is large enough the bosons may condense
into the same quantum state and a pion laser could be created~\cite{plaser}.
Similarly to this process, when a large number of bosonic atoms
are collected in a magnetic trap and cooled down to increase their
density in phase-space, the bosonic nature of the atoms reveals itself
in the formation of a Bose-Einstein condensate~\cite{atom} which can
be considered as a macroscopic quantum state. This condensation mechanism
may provide the key to the formation of atomic lasers in condensed matter
physics.

Thus it is essential to understand how the quantum statistical properties
of a dilute bosonic gas change if the density of the quanta increases.
In terms of multi-particle wave packet states this corresponds to the
increase of the overlap of the single-particle wave-packets and this is
indeed the key mechanism in our picture to provide the Bose-Einstein
condensation effects.
Due to the fact that the quanta in high energy physics are created
 by well localized sources, we would like to avoid the plane-wave
 approximation and utilize wave-packets
instead. In case of the Bose condensation of atomic vapors, the presence
of the magnetic trap prohibits the application of the outgoing plane-waves.
We hope that the wave packet formalism, presented herewith, may have
certain applications in that area of condensed matter physics as well,
due to the finiteness of the wave-packet sources. However,
we will not explore this possibility in details herewith, since
our main motivation for considering the problem of quantum statistical
correlations for many hundreds of overlapping wave-packets stems
from the experimental program in high energy heavy ion physics,
which itself has been motivated by a search for a new phase of the
strongly interacting matter. This new phase is referred to as
the Quark-Gluon Plasma (QGP), where  the quark and gluon
constituents of strongly interacting particles are expected to be
liberated from the confining forces of the strong interaction.
See refs.~\cite{qm96,s96} for recent reviews on the experimental and
theoretical developments in the quest for the QGP.

Bose - Einstein correlations  have already been
investigated in a large number of papers for various reactions of
high energy  physics (see for example the list
in refs.~\cite{bengt,bill,uli_sum}).
In particular, some models assume that pions are created at a given point
in space with prescribed momenta
that can only be approximately valid in a quantal description.
Such models correspond to  certain semi-classical Monte Carlo
simulations of high energy reactions, e.g.
\cite{spacer,venus,attila,jetset}. The Bose-Einstein correlations
are switched on only after the particle production is completed in refs.
{}~\cite{spacer,venus,RQMD} with the help of a weighting method that utilizes
the Wigner-function formalism~\cite{wigner,rel_wigner,pratt_csorgo}.
In the latter, the production of the pions
is given in space-time as well as in momentum space, since the
hadronic string model associates a  momentum space fragmentation with a
space-time fragmentation, see ref.~\cite{lundm}.  There are different
approaches incorporating both quantum mechanics and source models
for pions, such as the Wigner function formalism, developed by
Pratt~\cite{wigner}, and its different generalizations, for example
 the relativistic Wigner function formalism of ref.~\cite{rel_wigner}.

In the present paper we will utilize wave packets in order
 to resolve the clear contradiction
of the classical  assumptions in some models of
Bose-Einstein correlations with the tenets of quantum mechanics.
Although certain properties of the wave-packets have been already
utilized in previous
publications~\cite{rel_wigner,heinz_chap,wilk},
these were usually applied on a phenomenological level, utilizing e.g.
minimal wave packets on the single particle level to mimic the effects
of the uncertainty relations. However, in our case, we shall consider the
problems arising from the overlap between different wave-packets,
we shall discuss, how a Poisson multiplicity distribution is gradually
transformed to a multiplicity distribution  of
a Bose-Einstein condensate for a suitable choice of the wave packet size.
	({\it Note that this type of Bose-Einstein condensation is
	different from the text-book example of Bose-Einstein condensation
	in statistical mechanics, where the zero-momentum plane-wave
	modes are populated with a macroscopic amount of quanta.
	Here the wave-packet state with zero mean momentum will be
	populated with a very large number of particles}. Due to the
	optical coherence of the bosons in the zero mean momentum wave-packet
	state, this system is sometimes referred to as a laser, c.f.
	the term ``pion-laser" introduced by S. Pratt in ref.~\cite{plaser}).

We shall also see how the singe-particle spectrum and
the two-particle correlation functions  shall be
drastically changed if the overlap of the multi-particle
wave-packets reaches the critical size necessary to create the
Bose-Einstein condensate.

In high energy heavy ion reactions the system at break-up
is considered as a pion source, which in general can only be
described by a density matrix.
One may describe a density operator by a weighted sum of
projection operators. We will follow this method, and specify
the various pion sources in two steps.
In section ~\ref{s:rho} we summarize the generic formulae of
the density matrix formalism and
summarize the definitions of the physical quantities
under investigation. At this level,
projectors to $n$-particle wave-packet states shall be introduced.
These shall be constructed from single-particle wave-packet
states in subsection~\ref{ss:1wave}.
In subsection~\ref{ss:mwave} we specify a density matrix
that includes induced emission  on the level of multi-particle
wave-packets. We reformulate the multi-particle wave-packet density
matrix model in terms of new variables in such a manner that
an effective multi-particle plane-wave model is obtained,
with source parameters explicitly depending on the wave-packet size.
We utilize of the ring-algebra of multi-particle symmetrization
-- discovered by S. Pratt~\cite{plaser}
and described in details in ref.~\cite{chao} -- to reduce the
effective plane-wave model to a set of recurrence relations in
subsection~\ref{ss:recurrence}. We solve these recurrences
analytically:   A subset of the recurrence equations 
is solved analytically
in the case of arbitrary boson density in section~\ref{s:4} .
Complete analytic solutions are obtained for the
generating function of the multiplicity distribution.
Analytic results for
the single particle inclusive momentum distributions and
the two-particle inclusive correlation functions are presented
in subsections ~\ref{ss:rare} and ~\ref{ss:cond} for
a rare gas limiting case  and a highly condensed Bose-gas limit,
respectively.
 The onset of the wave-packet version of Bose-Einstein condensation is
studied analytically in section ~\ref{ss:onset}.
The analytical results presented in these sections
are new results not only for the multi-particle
wave-packet systems but also for the multi-particle symmetrization
of plane-wave systems.  In section~\ref{s:numerics}
we present numerically evaluated single-particle spectra
and two-particle correlation functions for the generic case.
Finally we summarize and conclude.

\section{Formulae for a generic case \label{s:rho}}
The density matrix of a generic quantum mechanical
system is prescribed as a sum of
density matrixes with different number of bosons,
\be
\hat \rho = \sum_{n=0}^{\infty} \, {p}_n \, \hat \rho_n,
\ee
the density matrix of the whole system is normalized as
\be
 {\rm Tr} \, \lef \hat \rho \ri = 1,
\ee
similarly to the density matrix characterizing systems with
a fixed particle number $n$,
which satisfy the normalization condition
\be
{\rm Tr} \, \lef \hat \rho_n \ri = 1,
\ee
from which it follows that
\be
 \sum_{n=0}^{\infty} {p}_n = 1.
\ee
The multiplicity distribution is prescribed in general by
the set of $\left\{ {p}_n\right\}_{n=0}^{\infty}$.
Later on we shall utilize the Poissonian multiplicity
distribution for the case when multi-particle symmetrization effects
can be neglected.

The density matrix describing a system with a fixed number
of bosons is given by
\be
\hat \rho_n = \int d\alpha_1 ... d\alpha_n
\,\,\rho_n(\alpha_1,...,\alpha_n)
\,\ket{\alpha_1,...,\alpha_n} \bra{\alpha_1,...,\alpha_n},
\label{e:rho_n}
\ee
where the states $\ket{\alpha_1,...,\alpha_n}$ denote $n$-particle
wave-packet boson states, which shall be prescribed in more details
in the next subsection.  The parameters $\alpha_i$ characterize a
single-particle wave-packet, and the multi-particle wave-packet states
are properly normalized to unity:
\be
\brak{\alpha_1,...,\alpha_n}{\alpha_1,...,\alpha_n} = 1.
	\label{e:ket_norm}
\ee
Thus  the normalization condition for the $n$-particle
density matrix can alternatively be written as
\be
\int d\alpha_1 ... d\alpha_n \,\,\rho_n(\alpha_1,...,\alpha_n) = 1.
	\label{e:w_norm}
\ee

The expectation value of an observable, represented by an operator
$\hat O$ is given by
\be
< \hat O > = {\rm Tr} \, \lef \hat \rho \, \hat O \, \ri
\ee

 The $i$-particle inclusive number distributions are given by
 \be
 N_i({{\bk}}_1,...,{\bk}_i) = {\bf Tr} \, \lef \, \hat\rho \,
        a^{\dag}({\bk}_1) ... a^{\dag}({\bk}_i) a({\bk}_i) ... a({\bk}_1) \,
\ri
 \ee
 which is normalized to the $i$-th factorial moment of the
 momentum distribution as
 \be
 \int d^3 {\bk}_1 \, ... \, d^3 {\bk}_i \,  N_i({\bk}_1,...,{\bk}_i)   =
 \langle n (n-1) ... (n-i+1) \rangle.
 \ee
 These inclusive number distributions can be built up from the exclusive
 number distributions of $N_i^m({\bk}_1,...,{\bk}_m) $ for all $m = i, i+1,
 ... $ describing the $i$-particle number distribution for fixed $m$
 multiplicity events as
 \be
 N_i({\bk}_1,...,{\bk}_i) =
	 \sum_{m=i}^{\infty} p_i N_i^{(m)}({\bk}_1,...,{\bk}_i)
	\label{e:d.n.i}
 \ee
 where
 \be
 N_i^{(m)}({\bk}_1,...,{\bk}_i) = {\bf Tr} \, \lef \, \hat\rho_m \,
         a^{\dag}({\bk}_1) ... a^{\dag}({\bk}_i) a({\bk}_i) ... a({\bk}_1) \ri
         \label{e:d1}
 \ee
 and
 \be
 \int d^3 {\bk}_1 \, ... \, d^3 {\bk}_i \, N_i^{(m)} ({\bk}_1,...,{\bk}_i) = m
... (m - i + 1)
 \label{e:d2}
 \ee

        The $i$-particle probability distributions
        are defined as

\be
 P_i({\bk}_1,...,{\bk}_i) = {\dst 1 \ov \langle n (n-1) ... (n-i+1) \rangle}
                N_i({\bk}_1,...,{\bk}_i)
	\label{e:d.p.i}
\ee
which are normalized to unity.
 For a fixed multiplicity of $m$ the $i$-particle probability distributions
 are defined as
 \be
 P_i^{(m)}({\bk}_1,...,{\bk}_i) = {\dst 1 \ov m (m-1) ... (m-i+1)}
                N_i^{(m)}({\bk}_1,...,{\bk}_i)
	\label{e:d.p.x}
 \ee
 which are also normalized to unity.
 The $i$-particle inclusive correlation functions are defined as
 \be
 C_i({\bk}_1,...,{\bk}_i) = {\dst P_i({\bk}_1,...,{\bk}_i) \ov \prod_{j=1}^i
P_1({\bk}_j)}
	\label{e:d.i.c}
 \ee
 while for a fixed multiplicity $m$ the
 $i$-particle exclusive correlation functions are defined as
\be
  C_i^{(m)}({\bk}_1,...,{\bk}_i) = {\dst P_i^{(m)}({\bk}_1,...,{\bk}_i) \ov
        \prod_{j=1}^i P_1^{(m)}({\bk}_j)}
	\label{e:d.x.c}
\ee

Let us note, that another definition of the $i$-particle (exclusive/inclusive)
correlation functions has been discussed recently in ref.~\cite{cnorm},
which will be denoted by $C_i^{N,(m)}({\bk}_1,...,{\bk}_i)$ and
$C_i^{N}({\bk}_1,...,{\bk}_i)$, respectively. These are defined as
\bea
  C_i^{N,(m)} ({\bk}_1,...,{\bk}_i) & = &
	{\dst N_i^{(m)}({\bk}_1,...,{\bk}_i) \ov
	\prod_{j=1}^i N_1^{(m)}({\bk}_j)},
  \label{e:d.x.cn} \\
 C_i^N({\bk}_1,...,{\bk}_i) & = &
	{\dst N_i({\bk}_1,...,{\bk}_i) \ov
		\prod_{j=1}^i N_1({\bk}_j)},
  \label{e:d.i.cn}
\eea
 and they are related to the probability correlations by a normalization
 factor,
\bea
	C_i^{N,(m)} ({\bk}_1,...,{\bk}_i) & = &
		{\dst m(m-1) ... (m-i+1) \ov m^i}
		C_i^{(m)} ({\bk}_1,...,{\bk}_i)
		\label{e:c.x.cn}\\
	C_i^{N}({\bk}_1,...,{\bk}_i) & = &
		{\dst \langle  m(m-1) ... (m-i+1) \rangle
		\ov \langle m \rangle^i }
		C_i ({\bk}_1,...,{\bk}_i).
		\label{e:c.i.cn}
\eea
	These relations indicate that for large values of $m$ the
	difference between the exclusive correlation functions  of
	eqs.~(\ref{e:d.x.cn}) and (\ref{e:d.x.c}) is small,
	but the definition may matter much more when the inclusive
	correlation functions are evaluated from eqs. ~(\ref{e:d.i.cn})
	or (\ref{e:d.i.c}).

The one- and two pion distributions are calculated as
\bea
N_2({\bk}_1,{\bk}_2) &= & {\bf Tr} \, \lef \, \hat \rho \, a^{\dag} ({\bk}_1)
a^{\dag} ({\bk}_2) a({\bk}_2) a({\bk}_1) \,\ri \cr
N_2({\bk}_1,{\bk}_2) &= & \sum_n \, {p}_n \, N_2^{(n)}({\bk}_1,{\bk}_2) \cr
N^{(n)}_2({\bk}_1,{\bk}_2) &= & {\bf Tr} \, \lef \, \hat \rho_n \, a^{\dag}
({\bk}_1)\, a^{\dag} ({\bk}_2) \,a({\bk}_2) \,a({\bk}_1)\, \ri
\eea
Similarly,
\bea
N_1({\bk}_1) &= & \sum_n \, {p}_n \, N_1^{(n)}({\bk}_1) \cr
N_1^{(n)}({\bk}_1) & = & {\bf Tr} \,\lef \, \hat \rho_n \,a^{\dag}({\bk}_1)
\, a({\bk}_1) \, \ri
\eea
We also have the following general relations
\bea
\int d^3 {\bk}_2 \, N^{(n)}_2({\bk}_1,{\bk}_2) &= & (n-1)\,
                 N_1^{(n)}({\bk}_1) \cr \label{e:norm12}
\int d^3 {\bk}_1 \, N_1^{(n)}({\bk}_1) & = & n \cr \label{e:norm1}
\int d^3 {\bk}_1 \, d^3 {\bk}_2 \,
	N_2({\bk}_1,{\bk}_2) &= & <n(n-1)>  \cr \label{e:exp12}
\int d^3 {\bk}_1 \, N_1({\bk}_1) & = & <n> \label{e:exp1}
\eea
The two-particle inclusive
Bose-Einstein correlation function is then given by the expression
\be
C({\bk}_1,{\bk}_2) =
C_2({\bk}_1,{\bk}_2) =
 { <n>^2 \over <n(n-1)> } \, {N_2({\bk}_1,{\bk}_2)
	\over N_1({\bk}_1) N_1({\bk}_2) }
        = {P_2({\bk}_1,{\bk}_2) \over P_1({\bk}_1) P({\bk}_2) }
\ee

The general considerations presented above
provide the framework for including the
effects arising from multi-particle Bose-Einstein
symmetrization using wave packets.
Before specifying how the $n$-particle density
matrixes $\hat \rho_n$ are related to the
$n$-particle wave-packets,
let us first introduce single-particle wave-packet states.
The simplest case,
when  a single particle is emitted from  a single source,
shall be specified in  subsection~\ref{ss:1wave}
and shall be extended for many particle wave-packet states emitted from
many sources in subsection~\ref{ss:mwave}.
We end this subsection
by specifying a density matrix for which
one can overcome analytically  the difficulty related to
the over-completeness of single-particle wave-packet states
which shall complicate the normalization of multi-particle
wave-packet states.
The higher-order symmetrizations are reduced to an
equivalent plane-wave problem in section~\ref{s:3},
that is solved analytically in section~\ref{s:4} in
certain limiting cases.  Numerical results are shown in the subsequent
section~\ref{s:numerics} to illustrate the new effects in the generic case.

\subsection{Single-particle wave-packets\label{ss:1wave}}

The building block for the definition of states
will be a  single particle wave packet,
which is described as follows.
  The operator $\hat { a}^{\dag} (\bx)$
  creates a boson (e.g. a pion) at point $\bx$.
This creation operator  $\axd $ can be decomposed as

{\begin{equation}
\axd =\int {d^3\bp \over (2 \pi)^{3\over 2}} \ \exp( i\bp \bx)
\apd.
\label{e:1}
\end{equation} }

The creation and annihilation operators in space and in momentum space
obey the standard commutation relations

\be
[\ax,\ \hat { a}^{\dag} (\bx^{\,\prime})] = \delta(\bx -
\bx^{\,\prime}) \label{e:2}
\ee

\be
[\ap,\ \hat  a^{\dag} (\bp^{\,\prime})] = \delta(\bp - \bp^{\,\prime})
\label{e:3}
\ee

Let us define a wave packet creation operator
as a linear combination of the creation operators in space,

{\begin{equation}
\alpha_i^{\dag} = \int {d^3\bx \over
(\pi / \sigma^2 )^{3\over 4} } \ \exp[-( \bx- \bxi_i)^{2} \sigma^{2}/2]\
        \exp[-i \bpi_i \bx] \ \axd .
\label{e:4}
\end{equation} }

Here $\alpha_i = (\bxi_i, \bpi_i, \sigma_i)$ refers to the parameters of the
wave packet: the center in space, in momentum space and the width
in momentum space, respectively. (The index $_i$ is introduced to
distinguish single-particle wave-packets with different parameters).
The wave-packet created by this operator is given by
\be
\ket{\alpha_i} = \alpha_i^{\dag} \ket0,
\label{e:5}
\ee
which is normalized properly as
\be
\brak{\alpha_i}{\alpha_i} = 1.
\label{e:6}
\ee
Although these wave packages are normalized to unity they are not
orthogonal; in fact they form an over-complete set~\cite{klauder}.
The commutator for the wave-packet creation and annihilation
operators is given by the overlap as:
\be
[\alpha_i,\alpha_j^{\dagger}] = \brak{\alpha_i}{\alpha_j},
\ee
as can be verified directly from the definition of the wave-packet
creation and annihilation operators.
The overlap between two wave-packets is given explicitly by
\begin{eqnarray}
\brak{\alpha_i}{\alpha_j} &= &
	\left( {\dst 2 \sigma_i \sigma_j \ov \sigma_i^2 +
	\sigma_j^2} \right)^{(3/2)}
	\exp\left(- {\dst (\sigma_j^2 \bxi_i - \sigma_i^2 \bxi_j)^2
		\ov 2 (\sigma_i^2 + \sigma_j^2)}
		- {\dst (\bpi_i - \bpi_j)^2 \ov 2 (\sigma_i^2 + \sigma_j^2)}
		\right)\times \cr
\null & \null &
		\exp\left( i (\bpi_i - \bpi_j)
		\left({\dst \sigma_i^2 \bxi_i + \sigma_j^2 \bxi_j
		\ov \sigma_i^2 + \sigma_j^2}\right) \right)
\end{eqnarray}

The coordinate space representation of a wave packet is given by
\be
\brak{x}{\alpha} = {1 \over (\pi / \sigma ^2) ^ {3 \over 4} } \
\exp[ - { (\bx-\bxi)^2 \sigma^2 \over 2} - i \bpi \bx ] \label{e:7}
\ee

Expansion ~(\ref{e:1})
can be used to obtain the momentum space representation as
\be
\brak{p}{\alpha} = {1 \over (\pi  \sigma ^2) ^ {3 \over 4} } \
\exp[ - { (\bp-\bpi)^2 \over 2 \sigma^2  } -
i \bxi (\bp - \bpi) ]
\label{e:8a}
\ee

Up to this point all quantities were defined at a given time
$t_0$, which may be considered as break-up time. The notation
$ \ket\alpha \equiv \ket{\alpha,t_0} $ reflects this fact.
 Assuming that after the break-up the wave packets
move without interaction, the time evolution of the wave packet
state is given as

\be
\ket{\alpha;t}  =
e^{iH(t-t_0)} \ \ket\alpha
 =
\int d^3 \bp \brak{p}{\alpha} \ \exp[ i \omega(\bp) ( t-t_0)] \ \apd \
\ket0
\label{e:9a}
\ee

  We introduce the notation
\be
w(\alpha,\bp) = \brak{p}{\alpha;t} = g(\bp, \bpi) \ \exp[ i
\omega(\bp) (t-t_0) - i (\bpi - \bp) \bxi ]
\label{e:10a}
\ee
with
\be
g(\bp, \bpi) = {1 \over [\sigma \pi]^{3\over
4}}\exp\bigl[-{(\bp - \bpi)^2 \over 2 \sigma ^2} \bigr].
\label{e:11a}
\ee
 The pure state in eq.~(\ref{e:7}) describes a wave packet with
mean momentum $\bpi_i$, mean position $\bx_i$ at time $\tau_i$, with
a spread of $\sigma = \Delta p$ in momentum space and a corresponding
$\sigma_x = \Delta R = {\hbar / \sigma}$ spread in configuration space.
Although
the location and the momentum of a pion
cannot be specified simultaneously  with an arbitrary precision,
the values of the parameters $\bpi,\bxi,\sigma$
are not subject to any quantum mechanical restriction.
For example, these values of the wave packet parameters
 can be
identified with the classical pion production points in the phase space,
as calculated from a Monte Carlo or a hydrodynamical model.

\subsection{Building multi-particle wave-packet states\label{ss:mwave}}

 The basic building block for the following
considerations is
 the operator $\alpha_i^{\dag}(t)$ which is defined by
\be
\alpha_i^{\dag}(t) = \alpha_i^{\dag} \ \exp [ i \omega(\bp) ( t-t_0) ]
\label{e:n7}
\ee
where the index $i$ refers to the different possible values of the
center of the wave packets in coordinate and momentum space.
For simplicity we may assume that all the wave packets are
emitted at the same instant.
Thus we indicate $(\bpi_i, \bxi_i, \sigma_i, t_0)$ by $\alpha_i$.
Later on when we evaluate the recurrence relations we shall
also assume that each wave packet has the same size,
$\sigma = \sigma_i$.

Let us assume that there are M boson sources in the system.
If one boson can be emitted from {\it any} of these
sources then the state vector can be written  as

{\begin{equation}
\ket{ \{ \alpha \ \};\ 1 ;\ t}= \ket{\alpha_1,...,\alpha_M;\ 1 ;\
t} \propto \sum_{i=1}^M \ \alpha^{\dag}_i (t)\ \ket0 .
\label{e:8b}
 \end{equation} }

This state is a linear combination
 of one boson states and is  not normalized. The normalization constant is to
be
determined from the condition $ \brak{ \ \{ \alpha_i \};\ 1;\ t}
{ \ \{ \alpha_i \};\ 1;\ t} = 1$.

The  $n$-boson states can in turn be given as

{\begin{equation}
\ket{ \{ \alpha_i \ \};\ n ;\ t}= \ket{\alpha_1,...,\alpha_M;\ n
;\ t} \propto [\sum_{i=1} ^M \ \alpha^{\dag}_i (t)]^n \ \ket0 .
\label{e:9b}
\end{equation} }

In the usual treatment $M=n$.
For example, if
$M = n = 2$, the
two-particle state is specified as

{\begin{equation}
\ket{\alpha_1,\alpha_2;\ 2 \; \
t} \propto [\ \alpha^{\dag}_1 (t) + \alpha^{\dag}_2 (t)] \
[\ \alpha^{\dag}_1 (t) + \alpha^{\dag}_2 (t)]\ \ket0 .
\label{e:10b}
 \end{equation} }

When using this type of states one usually introduces a phase-averaging
method to get back the Bose -Einstein enhancement of pions at small
relative momenta~\cite{gyulassy}. A {\it different} possible
$n$-boson state is given by

{\begin{equation}
\ket{ \{ \alpha_i \};\ \{ M_i \}; \ n ;\ t}=
\ket{\alpha_1,...,\alpha_L; \  M_1,...,M_L ; \
n; \ t} \propto \prod_{i=1} ^L \ [ \ \alpha^{\dag}_i (t)]^{M_i} \
\ket0
. \label{e:11b}
 \end{equation} }

where
{\begin{equation}
M_1 + M_2 + ... + M_L = n,
\end{equation} }
and $\alpha_i \ne \alpha_j$ for $i \ne j$.

We restrict our investigation to this latter case only.
In eq.~(\ref{e:11b}) there are exactly $n$ creation operators acting
one after the other on the vacuum, and the identical creation operators
are grouped together. The most general case is that similar wave-packets
can be created repeatedly and that it is not necessary to create two similar
wave-packets one after the other, but a different wave packet can be
created in between. Because of this reason, the general form of
eq.~(\ref{e:11b}) can be written as
\be
\ket{\alpha_1,...,\alpha_n;t} \propto \prod_{i=1}^n  \alpha^{\dag}_i(t) \ket0
\label{e:11c}
\ee
where similar values are allowed for arbitrary subsets of
$\{\alpha_i\}_{i=1}^n $. As a special case,
it is possible that all bosons are created
in the same wave-packet state.

The norm of the state defined in eq.~(\ref{e:11c}) can be
calculated using the fact that
\be
\bra{0} \ a(\bp_1)\ ... \ a(\bp_n) \ a^{\dag}(\bp_n^{\prime}) \ ... \
a^{\dag}(\bp_1^{\prime}) \ket{0} =
	\sum_{\sigma^{(n)}} \prod_{i=1}^n \
		[a(\bp_i),a^{\dag}(\bp_{\sigma_i}^{\prime})] \, = \,
	\sum_{\sigma^{(n)}} \prod_{i=1}^n \ \delta ( \bp_{i} -
			\bp_{\sigma_i}^{\prime} )
\ee
where the summation over $\sigma^{(n)} $ indicates the summation over the
permutations of the first $n$ positive entire numbers
and $_{\sigma_i}$ denotes the permuted value of $i$ in a given
permutation.
( Note that the subscript $_{\sigma_i}$
which stands for the permuted value of $i$ should not be confused
with the normal-size $\sigma_i$, which stands for the width
of the $i$-th wave packet in the momentum space.)
Similarly to the plane wave case, one has
\be
\bra{0} \ \alpha_1 \ ... \ \alpha_n \ \alpha_n^{\dag} \ ...\ \alpha_1^{\dag}\
\ket{0} =
	\sum_{\sigma^{(n)}} \prod_{i=1}^n \ [\alpha_i,\alpha^{\dag}_{\sigma_i}]
	=  \sum_{\sigma^{(n)}}
\prod_{i=1}^n \ \brak{\alpha_i}{\alpha_{\sigma_i}}
\ee

The
$n$ boson states, normalized to unity, are thus given as
\be
\ket{\ \alpha_1, \ ...\ , \ \alpha_n} = \ket{\ \alpha_1, \ ... ,
 \ \alpha_n; \ t} =
 {1\over \sqrt{ \displaystyle{\strut
 \sum_{\sigma^{(n)} }
\prod_{i=1}^n
\brak{\alpha_i}{\alpha_{\sigma_i}}
 } } }
\  \alpha^{\dag}_n (t) \ ... \
\alpha_1^{\dag}(t) \ket{0}.
\label{e:expec2}
\ee
Let us introduce the notation
\be
\gamma_{i,j} = \brak{\alpha_i}{\alpha_{\sigma_i}}
\ee
which can be further simplified if the width of the wave-packets
are identical, $\sigma_i = \sigma_j$. This simplifying condition is
in principle not necessary, but we shall apply it in the
forthcoming because this assumption will simplify certain recurrence
relations. If all the wave packets have the same width, one has
{
\begin{eqnarray}
\gamma_{i,j}& =&  \int d^3 \bp w^*(\bp,\alpha_i) w( \bp, \alpha_j)
\nonumber \cr
\null & = & \exp( - {(\bpi_i - \bpi_j)^2 \over 4
\sigma^2} )
 \exp( - {(\bxi_i - \bxi_j)^2  \sigma^2 \over 4 } )  \times \nonumber \\
\null & \null &
 \exp( i \bpi_i \bxi_j - i \bpi_j \bxi_i)
 \exp( i (\bpi_i + \bpi_j) (\bxi_i - \bxi_j) /2 )
 \label{e:gamma}
\end{eqnarray}
}

Note, that $\gamma_{i,i} = 1$ and $\gamma_{i,j} = \gamma_{j,i}^*
$.
 The summation over the
permutations $\sigma^{(n)}$ can be decomposed as summing over
transpositions. Since any permutation in $\sigma^{(n)}$ can be
built up from the product of at most $n-1$ transpositions $(i,j)$,
the summation over the permutations can be written as a
sum over the partial sums where a partial sum contains all the
terms which belong to a class of a given $k$ with $ 0 \le k \le
n-1$, where $k$ is the minimal number of transpositions
necessary to build up a given permutation:
\be
 \sum_{\sigma^{(n)} }
\prod_{i=1}^n
 \gamma_{i,\sigma_i}   = 1 + \sum_{(i,j)} \mid \ \gamma_{i,j}
\mid^2 + [ \sum_{(i,j);(j,k)}\ \gamma_{i,j}\ \gamma_{j,k} \
\gamma_{k,i} + \sum_{(i,j);(k,l)} \mid \ \gamma_{i,j} \mid^2
\ \mid\ \gamma_{k,l} \mid^2 ] \ +\ ...
        \label{e:trans}
\ee
For $n = 2$ and $n=3$ the normalization factors are given,
respectively, by the inverse square root of the quantities
\be
1 + \mid \ \gamma_{1,2} \mid^2
\ee
\be
1\ + \ \mid \ \gamma_{1,2} \mid^2
\ + \ \mid \ \gamma_{2,3} \mid^2
\ + \ \mid \ \gamma_{1,3} \mid^2
\ + \ \gamma_{1,2} \ \gamma_{2,3}\ \gamma_{3,1}
\ + \ \gamma_{1,3} \ \gamma_{3,2}\ \gamma_{2,1}
\ee

In the next section we shall introduce a new type of density matrixes
using the above defined set of normalized wave-packet states.
Later on we shall provide the analytical solutions to these models.

\section{A New Type of Solvable Wave-Packet Density Matrix \label{s:3_0}}

In the formulae displayed up to now the normalization factor
always appears as e.g. in eq.~(\ref{e:trans}).
For a reaction like Pb + Pb at SPS in a single  central
 collision event one has approximately 500 - 600 ${\pi}^-$ .
 Thus for this case we have the the sum over all possible permutation of
 the 600 pions, and each term in the sum is the product of 600 functions,
 depending on 1200 vector parameters.
 And this expression is in the denominator. Later on, this normalization factor
 will be multiplied with the distribution function of these parameters,
 and than one should integrate over these parameters.
 It is clear, that it is practically impossible to perform this task.
 A very interesting numerical simulation has been reported recently about
 a very reasonable numerical approximation to evaluate the $n$-boson
 symmetrization effects with the help of a Monte-Carlo
 algorithm~\cite{cramer}. However, the efficiency of such algorithms
 decreases as $1/(n!)$ thus it is practically impossible at the moment
 to go beyond 5-th or 6-th order explicit symmetrizations.

There is one special density matrix, however, for which one can
overcome such a difficulty even in an explicit analytical manner.
Namely, if one assumes, that we have
a system, in which the emission probability  of a boson is increased if
there is an other emission in the vicinity.
This would be an effect similar to the induced emission.
Such an $n$-boson density matrix, eqs.~(\ref{e:rho_n}-\ref{e:w_norm}),
 may have the form:
\be
\rho_n(\alpha_1,...,\alpha_n) =  {\dst 1 \ov {\cal N}{(n)}}
\lef \prod_{i=1}^n \rho_1(\alpha_i) \ri \,
\lef\sum_{\sigma^{(n)}} \prod_{k=1}^n \,
\brak{\alpha_k}{\alpha_{\sigma_k}}
\ri
\label{e:dtrick}
\ee
The coefficient of proportionality, ${\cal N}{(n)}$,
 can be determined from the condition
that the density matrix is normalized to unity,
\be
{\cal N}{(n)} = \int \prod_{i=1}^n d\alpha_i \rho_1(\alpha_i)
        \sum_{\sigma^{(n)}} \prod_{k=1}^n \, \gamma_{k,\sigma_k} =
         \sum_{\sigma^{(n)}} \int \prod_{i=1}^n d\alpha_i \rho_1(\alpha_i)
	\brak{\alpha_k}{\alpha_{\sigma_k}}
\ee
The density matrix given by eq.~(\ref{e:dtrick})
corresponds to the expectation that the creation
of a boson has a larger probability in a state, which is already
filled by another boson. The above model not only describes such a
source but it makes  possible to continue the calculation as well,
given that the last term of the density matrix cancels
the normalization factor of the overlapping wave-packet states.

The induced emission, that is implicitly built in to the above
definition~(\ref{e:dtrick}) of this density matrix,
can be made much more transparent,
if one evaluates the ratio
$\rho_n(\alpha_1,...,\alpha_n) / [\prod_{i=1}^n \rho_1(\alpha_i) ]$
for two special cases: one when each of the $n$ particles are emitted with the
same wave-packet parameter $\alpha_i$ (maximal overlap) and
the other, when the overlap between the
wave packets of any pair from the $n$ can be negligible.
If $n$ particles are emitted in the same wave-packet state, one
has
\be
{\dst \rho_n(\alpha_1,...,\alpha_1)  \ov [\rho_1(\alpha_1)]^n }
 = {\dst n! \ov {\cal N}{(n)} },
\ee
while if the overlap between any of the wave packets is negligible,
we obtain
\be
{\dst \rho_n(\alpha_1,...,\alpha_n) \ov \prod_{j=1}^n \rho_1(\alpha_j)}
 = {\dst 1  \ov {\cal N}{(n)} } \qquad \mbox{\rm (no overlap)}.
 \ee
In general, the overlap of the wave-packets determines the
magnitude of the enhancement of the density matrix:
\be
{\dst \rho_n(\alpha_1,...,\alpha_n) \ov \prod_{j=1}^n \rho_1(\alpha_j)}
 = {\dst  \sum_{\sigma^{(n)}} \prod_{k=1}^n
	\brak{\alpha_k}{\alpha_{\sigma_k}}
	\ov {\cal N}{(n)} } \qquad \mbox{\rm (overlap)}.
 \ee
 Thus the density matrix given in eq.~(\ref{e:dtrick}) describes a
 quantum-mechanical wave-packet system with induced emission, and the
 amount of the induced emission is controlled by the overlap of the $n$
 wave-packets, yielding a weight in the range of $[1,n!]$.
 Although it is very difficult numerically to operate with such a
 wildly fluctuating weight, we shall show that the special form of
 our density matrix yields a set of recurrence relations 
 that can be evaluated numerically in an efficient manner.
 The essence of the method is the reduction of the problem to the
 already discovered ``ring" - algebra of permanents for plane-wave
 outgoing states~\cite{plaser}.

 We have a larger freedom
to choose the form of $\rho_1(\alpha)$, the density
matrix describing the distributions of the
parameters of wave packets. However, one can
perform the calculations analytically when choosing Gaussian forms.
For simplicity, we do not discuss the fluctuations of the wave-packet
sizes in the forthcoming, although a fluctuating wave-packet size
with Gaussian random distribution would lead to a similar mathematical
structure of the solution.

Thus for  the single-particle density
matrix we assume the following form:
\bea
\rho_1(\alpha)& = &\rho_x(\bxi)\, \rho_p (\bpi)\, \delta(t-t_0) \cr
\rho_x(\bxi) & = &{1 \over (2 \pi R^2)^{3/2} }\, \exp(-\bxi^2/(2
R^2) )\cr
\rho_p(\bpi) & = &{1 \over (2 \pi m T)^{3/2} }\, \exp(-\bpi^2/(2 m
T) )
\eea

Note, that this choice corresponds to a non-relativistic, non-expanding
static source at rest in the frame where the calculations are performed.

 This completes the specification of the model.
{\it If one is interested only in the numerical results,
we recommend to jump directly to Section \ref{s:numerics}}, where
multiplicity distributions, particle spectra and correlations
are plotted from a numerical evaluation of the model.
For those more theoretically interested, we present in  the next sections
the analytic solution of this  model. We may comment that these solutions
are not easily obtainable, however, the algebraic beauty of the
structure of the equations well compensates for the difficulty.

\section{\label{s:3} Algebraic Evaluation of the Model}
In this section we make algebraic manipulations that are
necessary to reduce the multi-particle wave-packet problem
to an analytically solvable plane-wave problem.

The $n$ particle distribution for the system containing $n$ bosons
is given by the expression:

\bea
 N^{(n)}_{n}\,  (\, {\bk}_1,{\bk}_2,...,{\bk}_n) & = &  { 1 \over \dst {\cal
N}{(n)} }
 {\displaystyle \int \prod_{m=1}^n d\alpha_m \,
\rho(\alpha_m) \sum_{\sigma^{(n)}}
 \prod_{i=1}^n   w^*({\bk}_i,\alpha_i)\,w({\bk}_i,\alpha_{\sigma_i})
}
\label{e:na2a}
\eea

Since the indices run through all possible values, we can
exchange the index of ${\bk}$ and $\alpha$ in the second $w$.

\bea
N^{(n)}_{n}\, (\, {\bk}_1,{\bk}_2,...,{\bk}_n) & = &  { 1 \over \dst {\cal
N}{(n)} }
 {\displaystyle \int \prod_{m=1}^n d\alpha_m \,
\rho(\alpha_m) \sum_{\sigma^{(n)}}
 \prod_{i=1}^n   w^*({\bk}_i,\alpha_i)\,w({\bk}_{\sigma_i},\alpha_i)
}
\label{e:na2b}
\eea

One more rewriting becomes possible by using the identity
$(\prod_{i=1}^n A_i) (\prod_{j=1}^n B_j) = \prod_{i=1}^n (A_i B_i)$:

\bea
 N^{(n)}_{n}\, ( \, {\bk}_1,{\bk}_2,...,{\bk}_n) & = &  { 1 \over \dst {\cal
N}{(n)} }
 {\displaystyle
 \sum_{\sigma^{(n)}}
 \prod_{i=1}^n  \int d\alpha_i \,  w^*({\bk}_i,\alpha_i)\,\rho(\alpha_i)\,
w({\bk}_{\sigma_i},\alpha_i) }
\label{e:na3}
\eea

Let us  introduce the auxiliary quantity
\be
\overline\rho({\bk}_i,{\bk}_j) = \int d\alpha_i \rho(\alpha_i)
w^*({\bk}_i,\alpha_i)
        w({\bk}_j,\alpha_i)
\ee
where the overline and the two arguments distinguish this
auxiliary quantity from the single-particle density matrix
$\rho(\alpha)$, introduced earlier.

Using these notations Eq.~(\ref{e:na3}) can be rewritten as:

\bea
N^{(n)}_{n} \, (\, {\bk}_1,{\bk}_2,...,{\bk}_n) & = & { 1 \over \dst {\cal
N}(n) }
 {\displaystyle
 \sum_{\sigma^{(n)}}
 \prod_{i=1}^n   \overline\rho({\bk}_i,{\bk}_j)
 }
\label{e:na4a}
\label{e:s.n.n}
\eea

With  this notation the ``source function" will  be defined as

\be
        S_n({\bk}_1,...,{\bk}_n) = \sum_{\sigma^{(n)}} \prod_{i=1}^n
                \overline\rho({\bk}_i,{\bk}_{{\sigma}_i})
	\label{e:s.s.n}
\ee

and further:

\bea
N^{(n)}_{n}\, (\, {\bk}_1,{\bk}_2,...,{\bk}_n) & = &   { 1 \over \dst {\cal
N}(n) }
  S_n({\bk}_1,...,{\bk}_n)
\label{e:na4b}
	\label{e:s.n.s}
\eea
Using Eq.~(\ref{e:na4a}) and the definitions ~(\ref{e:d1},\ref{e:d2}),
the one and the two particle distributions are cast in the form
 \be
        N_1^{(n)}({\bk}_1) = {n \over {\cal N}(n)}
        \int \prod_{l=2}^n d{\bk}_l S_n({\bk}_1,...,{\bk}_n)
\ee
 \be
        N_2^{(n)}({\bk}_1,{\bk}_2) = {n(n-1) \over {\cal N}(n)}
        \int \prod_{l=3}^n d{\bk}_l S_n({\bk}_1,...,{\bk}_n)
\ee
{}From this it is clear that the constant of normalization ${\cal N}(n)$
can be expressed as
\be
        {\cal N}(n) = \int \prod_{l=1}^n d{\bk}_l S_n({\bk}_1,...,{\bk}_n).
\ee
At this point one can realize, that the above equations have
algebraic structure similar to those discussed in
 Refs.~\cite{plaser,chao} , thus we can use their method
to perform the summation over the permutations.

Up to this point the expressions in this subsection refer to
events with fixed pion multiplicity.

For an average over many events, one has to treat the multiplicity
distributions too.

Following Refs.~\cite{plaser,chao} we introduce the auxiliary parameter, $n_0
$,
which correspond to the mean multiplicity of a source containing
classical particles (not symmetrized system). If the system is characterized
by a Boltzmann distribution, then this $n_0$ is a function of chemical potential,
the temperature and the volume (characterized by the radius, R. )

Let us assume a Poissonian multiplicity distribution for the case when
the Bose-Einstein effects are switched off (denoted by $p_n^{(0)}$), i.e.
let us assume that
\be
        {p}^{(0)}_n = {n_0 ^n \over n!} \exp(-n_0),
\ee

Let us define
\be
        G_1(p,q) = {n_0 \over 1!} \overline\rho(\bp,\bq)
\ee
\be
        G_2(p,q) = {n_0^2 \over 2!}
                \int \overline\rho(\bp,\bp_1) d\bp_1
                \overline\rho(\bp_1,\bq)
\ee
\be
        G_i(p,q) = {n_0^i \over i!}
                \int
                \overline\rho(\bp,\bp_1) d\bp_1
                \overline\rho(\bp_1,\bp_2) ...
                \overline\rho(\bp_{i-2},\bp_{i-1}) d\bp_{i-1}
                \overline\rho(\bp_{i-1},\bq)
\ee
With these definitions
we find that the case of multi-particle wave packets
can be considered as a formally equivalent
plane -wave system due to the complete formal analogy between
the above equations  and those of ref.~\cite{chao} .

The above mentioned problem was shown to be solvable with the help
of the so-called ring algebra discovered first
by S. Pratt~\cite{plaser}. Utilizing that algebra,
a set of recurrence relations were obtained  that contained
some effective parameters of the $T$ matrix which were
assumed to have a Gaussian form. In our case, the width of the single-pion
wave packet  and the width of the distributions of the wave-packet
centers are taken into account explicitly, thus a wave-packet size
appears as one of the parameters of the recurrence given below.
We shall also present certain analytic solutions to these recurrence relations
and investigate them numerically as well.

\subsection{Recurrence relations\label{ss:recurrence}}
Having performed the reduction of the multi-particle wave-packet
problem to the multi-particle plane-wave problem, we now can utilize the
reduction of the multi-particle plane-wave problem
to a set of recurrence relations that were given first
in refs.~\cite{chao,plaser}.

The multiplicity distribution
(with the inclusion of the symmetrization effects),
the one- and two-particle distribution for a system containing
$n$ pions can be expressed with the help of the
three definitions that relate
these observables to the elements of the recurrence relations, four
recurrence relations and four initial conditions (one for each recurrence
relation). Two further equations are necessary to define certain auxiliary
variables that are related to the recurrence relations and the observables.
 Thus one ends up a set of 14 equations, as given below.
The observables are defined as
\be
	{p}_n  =  {\dst \omega_{n} \ov \sum_{k=0}^{\infty} \omega_{k} }
			\label{e:3.0}
			\label{e:d.1}
\ee
\be
	P^{(n)}_1(\bk)  =  {\dst 1
		 \ov n \omega_{n}}
                \sum_{i=1}^n  \omega_{n-i} G_i (\bk,\bk)
		,
			\label{e:3.1}
			\label{e:d.2}
\ee
\be
	P^{(n)}_2({\bk}_1,{\bk}_2)  =
		{\dst 1
		 \ov n(n-1) \omega_n}
		\sum_{l=2}^n
		\sum_{m=1}^{l-1}
		\omega_{n-l} \left[
			G_m({\bk}_1,{\bk}_1) G_{l-m} ({\bk}_2,{\bk}_2)
                + G_m({\bk}_1,{\bk}_2) G_{l-m}({\bk}_2,{\bk}_1)\right]
		,
		\label{e:3.2}
			\label{e:d.3}
\ee
in order to relate these definitions to the recurrence relations, one
introduces two auxiliary quantities as
\bea
        G_n(\bp,\bq) & = &
		n_0^n \, h_n \exp(-a_n (\bp^2 + \bq^2) + g_n \bp \bq),
                        \label{e:3.7}
			\label{e:a.1} \\
        C_n & = & {\dst 1 \ov n} \int d^3 p G_n(p,p), \label{e:3.4}
             \,\, = \,\, h_n {\dst n_0^n \ov n }
		\left( {\dst \pi \ov{2 a_n -g_n}}\right)^{3/2}.
		\label{e:3.5a}
		\label{e:a.2}
\eea
The physical interpretation of the quantities $\omega_n $ and $C_n$
shall be discussed later. The recurrence relations correspond to the solution
of the ring-algebra~\cite{chao,plaser} are given as
\bea
        \omega_n & = & {\dst 1 \ov n}
		\sum_{l=1}^n l C_l \omega_{n-l}
		\label{e:r.1}
		\label{e:3.3} \\
        h_{n+1}& = & h_1 h_n
		{\dst \pi^{3/2} \ov (a_1 + a_n)^{3/2}}
                \label{e:3.8}
		\label{e:r.2} \\
        a_{n+1} & = & a_1 - {\dst g_1^2 \ov 4 (a_1 + a_n)}
		\label{e:3.9}
		\label{e:r.3} \\
        g_{n+1} & = & {\dst g_1 g_n \ov 2(a_1 + a_n)}
		\label{e:3.10}
		\label{e:r.4}
\eea
The initial conditions or the starting elements of the recurrence are
\bea
        \omega_0 & = & 1,
		\label{e:3.6}
		\label{e:i.1} \\
        h_1 & = & {\dst 1 \ov [\pi \sigma_T^2]^{3/2}} ,
		\label{e:3.11}
		\label{e:i.2} \\
        a_1 & = & {\dst 1 \ov 2 \sigma_T^2} + {\dst R_{eff}^2 \ov 2},
                \label{e:3.12}
		\label{e:i.3}
		\\
        g_1 & = & R_{eff}^2,
		\label{e:3.13}
		\label{e:i.4}
\eea
which yield the following value for $C_1$ :
\bea
        C_1 & = & n_0.
		\label{e:3.5b}
		\label{e:i.5}
\eea
In the initial conditions eq. (\ref{e:i.2}-\ref{e:i.4}) the
following notation is introduced:
\bea
        \sigma_T^2 & = & \sigma^2 + 2 m T, \label{e:3.14} \label{e:sigt} \\
        R_{eff}^2 & = & R^2 + {\dst m T \ov \sigma^2 \sigma_T^2}.
					\label{e:3.15} \label{e:reff}
\eea
or using the spatial spread, $\sigma_x$,  and the effective
temperature $T_{eff}$, that
characterizes the system
before symmetrization effects are taken into account,
one can write
\bea
        T_{eff} & = &  T + \sigma^2/ (2 m) = T + 1/(2 m \sigma_x^2),
			\label{e:3.16} \\
        R_{eff}^2 & = & R^2 + {\dst \sigma_x^2\ov 2}  {\dst T \ov  T_{eff}}.
\label{e:3.17}
	\label{e:teff}
\eea
Observe that the above recurrence relations correspond to the
pion laser model of S. Pratt when a replacement $ R \rightarrow R_{eff}$
and $T \rightarrow T_{eff}$ is performed . The parameters of the $T$ matrix
in S. Pratt's model are thus interpreted here as effective parameters and
expressed with the help of the radius of the
source of the wave-packets, the temperature of the source of the wave-packets
and the width of the wave-packets in our multi-particle wave-packet model.

Thus the reduction of the multi-particle wave-packet model to an equivalent
plane-wave model is completed. These recurrence relations were studied
numerically but analytic solutions to these equations were not given before.
Although numerical investigation of the presented
recurrence relations were reported e.g. in refs.~\cite{plaser,chao},
to apply these kind of numerical investigations
 for multiplicities of few 1000 high numerical
precision and special technique is needed to treat the factorials
of large numbers appearing in the algorithm. We present some new numerical
results in this multiplicity range in Section~\ref{s:numerics}.

Before proceeding to the solution of the above
recurrences, let us note,
that in eqs.~(\ref{e:d1}-\ref{e:d.3}) the multiplicity
distribution and the {\it exclusive} single-particle and two-particle
momentum distributions were given only, i.e. the particle spectra that are
measured, which belong to special events with fixed multiplicity $n$.
Such events occur with a probability of $p_n$, i.e. they constitute a
sub-set of all the events. To get the more readily observable
{\it inclusive} momentum distributions -- hence the inclusive
correlation functions -- one  has to average over all the multiplicities
with weight factors $p_n$. To our best knowledge, such an averaging
of the exclusive momentum distributions was not performed before
in similar kind of models, neither analytically nor even numerically,
due to numerical difficulties involved in evaluating the
exclusive distributions for a large number of $n$.
However, the following analytical
considerations can be utilized  to simplify this task.
Let us consider the identities
\bea
\left( \sum_{j=0}^{\infty} a_j \right)
\left( \sum_{k=1}^{\infty} b_k \right)  & = &
	\sum_{n=1}^{\infty} \left( \sum_{m=1}^n a_{n-m} b_m \right)
	\label{e:dsum} \\
\left( \sum_{j=0}^{\infty} a_j \right)
\left( \sum_{k=1}^{\infty} b_k \right)
\left( \sum_{l=1}^{\infty} c_l \right)
	& = &
	\sum_{n=2}^{\infty} \left( \sum_{m=2}^n
	\sum_{r=1}^{m-1} a_{n-m} b_{m-r} c_r \right)
	\label{e:tsum}
\eea
When inserting the expressions for the exclusive momentum distributions
as given by eqs.~(\ref{e:d.1} - \ref{e:d.3}) to the definition of
the inclusive momentum distributions in terms of the exclusive ones,
eqs.~(\ref{e:d.n.i}-\ref{e:d.p.x}), exactly this type of products of
infinite sums appear, in the form of the r.h.s. of the above relations.
When rewriting these in terms of the l.h.s. of the identities,
one of the factors shall be $\sum_{n = 0}^{\infty} p_n = 1$ that 
simplifies~\cite{plaser} the result as
\bea
	N_1(\bk) & = &
		 \sum_{j = 1}^{\infty} G_j(\bk,\bk),
	\label{e:d.i.n1}
\eea
\bea
	N_2(\bk_1,\bk_2) & = &
		 \left(\sum_{i = 1}^{\infty} G_i(\bk_1,\bk_1) \right)
		 \left(\sum_{j = 1}^{\infty} G_j(\bk_2,\bk_2) \right)
		+
		 \left(\sum_{i = 1}^{\infty} G_i(\bk_1,\bk_2) \right)
		 \left(\sum_{j = 1}^{\infty} G_j(\bk_2,\bk_1) \right).
	\label{e:d.i.n2}
\eea
These simple and beautiful relations will be useful when evaluating
the inclusive number distributions analytically as well as numerically.
The probability distributions can be obtained from eqs.
{}~(\ref{e:d.p.i}).

The importance of the proper normalization of the
two-particle inclusive correlation functions was emphasized recently in
ref.~\cite{cnorm}. We are now in the position that a formula of
general validity can be {\it derived} that involves no approximation
and can be obtained without any reference to the details of the
source model that determines the functions $G_i(\bk_1,\bk_2)$.
As long as the $n$-particle exclusive number-distributions have the form
of eqs.~(\ref{e:s.n.n} - \ref{e:s.n.s}), the
probability distributions are given by the
functional equations of eq.~(\ref{e:d.1} - \ref{e:d.3})
and the above eqs. ~(\ref{e:d.i.n1} - \ref{e:d.i.n2}) follow.
In such cases it is possible to introduce the auxiliary quantity
\be
	G(\bk_1,\bk_2) = \sum_{i = 1}^{\infty} G_i(\bk_1,\bk_2).
	\label{e:d.ggg}
\ee
(In the Wigner-function formalism, this quantity
corresponds to the on-shell Fourier-transform of the emission function
$S(x,\bk)$.) With this notation, both kind of two-particle inclusive
correlation functions, eqs.~(\ref{e:d.i.c},\ref{e:d.i.cn}) can be evaluated
as
\bea
	C_2^N(\bk_1,\bk_2)  & = & 1 +
		{ \dst G(\bk_1,\bk_2) G(\bk_2,\bk_1) \ov
		G(\bk_1,\bk_1) G(\bk_2,\bk_2) },\\
	C_2(\bk_1,\bk_2)  & = &
		{\dst \langle n \rangle^2 \ov \langle n(n-1)\rangle }
		\left( 1 +
		{ \dst G(\bk_1,\bk_2) G(\bk_2,\bk_1) \ov
                G(\bk_1,\bk_1) G(\bk_2,\bk_2) }
		\right).
\eea
Thus we find, in agreement with ref.~\cite{cnorm}, that in general
a non-trivial pre-factor appears before the two-particle inclusive
momentum distribution $C_2(\bk_1,\bk_2)$, that can be transformed away
if the correlation function is defined as the ratio of the number of
counts, eq.~(\ref{e:d.i.cn}). Alternatively, this result can be interpreted
in another manner: when measuring the inclusive correlation function
as the ratio of the detection probabilities, eq.~(\ref{e:d.i.c}),
an overall normalization constant has to be introduced that will be
not arbitrary but will have to  be equal to
$\langle n\rangle^2/\langle n(n-1) \rangle$,
the inverse of the second normalized factorial moment of the
multiplicity distribution.

In the next sub-section we re-write these recurrence relations into
dimensionless forms and solve analytically three of them completely
and present a formal solution for the fourth recurrence too.
We present the leading order complete solution of the model both in the
rare gas and in the very dense Bose-gas limiting cases.

\section{Analytic Solutions \label{s:4}}

 When one assumes a Gaussian form for the wave-packets and for the
distribution functions of the parameters of the wave-packets,
the functional equations for $G_n$-s are reduced to the set of
recurrence relationships presented in the previous section.
The recurrence equations for the exclusive distributions
 were known since their discovery in 1993 (except for the
 inclusion of wave-packet sizes into the definition of $R_{eff}$ and
 $T_{eff}$). Thus, a number of properties of these recurrences were
 explored but as far as we know, only numerically.
 In particular, it has been observed that at a critical density
 $n_c$ the stimulated emission of bosons over-compensates for the
 decrease of the unsymmetrized $p^{(0)}_n$ probabilities and a lasing
 effect or coherent behavior appears. The condensation is characterized
 by the divergence of $p_n$ (symmetrized)  probabilities
 with increasing values of $n$, by an appearance of a low-momentum
 peak in the single-particle spectrum and by a decreasing
 intercept of the correlation function: $C({\bk},{\bk}) < 2$.
 However, no analytical results were
 known about the spectrum and the correlations
 at the condensation point as well as in other limiting cases,
 only the exclusive correlation functions were evaluated numerically
 but the evaluation of the inclusive correlations was not performed.

 Note also, that until now $n_0$ was just the
 mean number of  bosons before the symmetrization, and
 the parameters $n_0$, $R_{eff}$ and $T_{eff}$ were just interpreted
 in terms of the theoretical input values of the correlated system.
 It seemed that the actual single-particle spectra and correlation
 function can only be determined by numerically solving the  recurrence
 relations. The invention of the recurrence relation in ref.~\cite{plaser}
 was already a huge step forward, since the number of steps that
 are required for the solution of the $n$-particle symmetrization
 in general increase as $n! \propto \exp[n (\ln n - 1)] $ which increases
 faster than any polynomial - or a non-polynomially (NP) hard problem.
 The NP-hard problems are very difficult and inefficient
 to handle numerically for large values of $n$.
 S. Pratt reduced this difficult, NP-hard
	case to a set of recurrence relations,
 where the number of necessary steps to evaluate the observables
 increases only as slowly as $n^2$ and thus the problem is solvable
 even for large values of $n \simeq 1000 $ within a few minutes on the
 current computers.

 Here, we would like to present the first analytical solution
for the multi-particle symmetrization of the wave-packets
--- no numerical evaluation of the recurrences will be necessary,
 and the time needed to compute the result will be independent of $n$.
 The solution of the problem is possible  in the rare and the dense
	gas limiting cases, while in the general case the multiplicity
 distribution shall be given in terms of its combinants.

 Before presenting the analytical solution of the recurrence equations,
 let us re-formulate the recurrences in eqs.~(\ref{e:r.2} - \ref{e:r.4})
 for new, dimensionless quantities. Note that eq.~(\ref{e:r.1}) is already
 referring to dimensionless quantities thus there is no need to reformulate it.

 Let us introduce the following dimensionless variables:
 \bea
	A_n  & = & \sigma_T^2 (a_1 + a_n) \label{e:di.a} \\
	H_n  & = & h_n / h_1 \, = \, h_n / (\pi \sigma_T^2)^{3/2}
		\label{e:di.h}\\
	G_n  & = & \sigma_T^2 g_n \label{e:di.g} \\
	x    & = & R_{eff}^2 \sigma_T^2 \label{e:di.x}
 \eea
 Note that the variables indicated by the upper-case $A_n$, $H_n$ and $G_n$
 are essentially the dimensionless versions of the variables $a_n$, $h_n$ and
 $g_n$. The quantity $x$ corresponds to a dimensionless measure of the
 phase-space available for a single quanta. Extremely rare gas corresponds to
 the limit $ x \rightarrow \infty$ while a very dense Bose-gas corresponds to
 the $ x \rightarrow 0$ limiting case. Note also that the dimensionless
 variable $x$ should not be confused with the vector ${\bf x}$ indicating
 a position in space.

 The recurrence relations ~(\ref{e:r.2} - \ref{e:r.4}) can be re-written
 for the dimensionless variables as
 \bea
	A_{n+1} & = & A_1 - {\dst x^2 \over 4} {1 \over A_n}  ,
			\label{e:dr.1}\\
	H_{n+1} & = & H_n {\dst 1 \over A_n},
			\label{e:dr.2}\\
	G_{n+1} & = & G_n {\dst x \over 2 A_n} .
			\label{e:dr.3}
 \eea
 The initial conditions for these recurrences read as
 \bea
	A_1 & = & 1 + x, \label{e:di.1}\\
	H_1 & = & 1, \label{e:di.2} \\
	G_1 & = & x. \label{e:di.3}
 \eea
  The dimensionless recurrences of eqs.~(\ref{e:dr.1}-\ref{e:dr.3})
 can be solved exactly with the help of the auxiliary quantity
 \bea
	Y_n & = & \prod_{i = 1}^n A_i, \label{e:a.y}
 \eea
	since the solution of eqs.~(\ref{e:dr.2}-\ref{e:dr.3}) is
	given as
 \bea
	H_{n+1} & = & {\dst 1 \ov Y_n^{(3/2)} } \label{e:dr.2s} \\
	G_{n+1} & = & {\dst x^{n+1} \ov 2^n Y_n } \label{e:dr.3s}
 \eea
	where the initial conditions of eqs.~(\ref{e:di.2}-\ref{e:di.3})
	are already taken into account. The remaining recurrence relation,
	eq.~(\ref{e:dr.1}) is easy to solve if one rewrites this as
 \bea
	Y_{n+1} = ( 1 + x ) Y_n - {\dst x^2 \ov 4} Y_{n-1}, \label{e:dr.1.1}
 \eea
	that can be further re-formulated for an arbitrary value of the
	parameter $\gamma \ne 1 + x$ as
 \bea
	Y_{n+1}  - \gamma Y_n & = & (1 + x - \gamma)
	\left[ Y_n - {\dst x^2 \ov 4 ( 1 + x -\gamma) } Y_{n-1} \right]
		\label{e:dr.1.2}
 \eea
	and this equations becomes solvable if the value of the parameter
	$\gamma $ is chosen such that
 \bea
	\gamma & = & {\dst x^2 \ov 4 ( 1 + x - \gamma) }.
 \eea
	This equation has the following two roots:
 \bea
	\gamma_{\pm}
		& =  &{\dst 1\ov 2} \left( 1 + x \pm \sqrt{1 + 2 x} \right)
			\label{e:gam.s}
 \eea
	Note that these roots satisfy the following useful algebraic
	relations:
 \bea
	\gamma_+ + \gamma_- & = & 1 + x, \\
	\gamma_+ \gamma_- & = & x^2 / 4, \\
	\gamma_+^{(1/2)} - \gamma_-^{(1/2)} & = & 1.
 \eea
	One may choose any of $\gamma_+$ or $\gamma_-$ to solve the third
	recurrence equation by re-scaling eq.~(\ref{e:dr.1.1}) and introducing
 \bea
	y_n^{\pm} & = & Y_{n+1} - \gamma_{\mp} Y_n
 \eea
	One finds that
 \bea
	y^{\pm}_n & = & \gamma_{\pm} y^{\pm}_{n-1}, \\
	y^{\pm}_1 & = & \gamma_{\pm}^2,
 \eea
	thus the solution for the third recurrence relation can be written as
 \bea
	y^{\pm}_n & = & \gamma_{\pm}^{n+1}, \label{e:dr.1s.y}
 \eea
	which yields the solution for $ Y_n$ as
 \bea
	Y_n & = & {\dst \gamma_{\pm}^{n+1} - \gamma_{\mp}^{n+1}
		\ov	 \gamma_{\pm} - \gamma_{\mp} } \label{e:dr.1s.Y}
 \eea
	that can be substituted back to eqs.~(\ref{e:dr.2s}-\ref{e:dr.3s})
	to get an explicit solution. Then the variables that have dimensions
	can be also calculated  easily, and their solution in terms of
	$x = R_{eff}^2 \sigma_T^2 $ is especially simple if the variables
	defined in eq.~(\ref{e:gam.s}) are used:
 \bea
	a_n & = & {\dst \gamma_+ - \gamma_- \over 2 \sigma_T^2}
			{\dst \gamma_+^n + \gamma_-^n \ov
			      \gamma_+^n - \gamma_-^n}
		\label{e:a.s} \\
	h_n & = & {\dst 1 \ov [ \pi \sigma_T^2]^{3/2} }
		\left[	{\dst  	\gamma_+ - \gamma_- \ov
			\gamma_+^n - \gamma_-^n }\right]^{(3/2)}
		\label{e:h.s}\\
	g_n & = & {\dst 2\ov \sigma_T^2 } (\gamma_+ - \gamma_-)
		{\dst (\gamma_+ \gamma_- )^{(n/2)} \ov
		\gamma_+^n - \gamma_-^n }
		\label{e:g.s}
 \eea
	and these solutions can be utilized to evaluate the quantity
	$C_n$ from the first part of eq.~(\ref{e:a.2}) as
 \bea
	 C_n & = & {\dst n_0^n \ov n}
			{\dst 1\ov \left[ \gamma_+^{(n/2)} -
				\gamma_-^{(n/2)} \right]^3 } \,
		= \, t^n
			{\dst 1\ov \left[ 1 - v^n \right]^3 },
		\label{e:c.s} \\
	t & = & {\dst n_0 \ov \gamma_+^{(3/2)} }, \\
	v & = & \sqrt{ {\dst \gamma_- \ov \gamma_+}}, \\
	0 & \le & v < 1.
 \eea

	Note at this point, that the behavior of $p_n$ for large values of
	$n$ is controlled by $n C_n$, a quantity that can have a kind
	of critical behavior in the present model. Namely,
	the large $n$ behavior of $n C_n $ depends on the ratio of
	$t = n_0 / \gamma_+^{(3/2)}$, since for large values of $n$,
	we always have $v^n = (\gamma_- / \gamma_+)^{(n/2)} << 1$.
	One may introduce a critical value of $n_0$, indicated as
 \bea
	n_c & = & \gamma_+^{(3/2)} = \left[ {\dst 1 + x + \sqrt{1 + 2 x}
			\ov 2 } \right]^{(3/2)}
 \eea
	and one may observe at this point that if $n_0 < n_c$,
	one has $lim_{n \rightarrow \infty} n C_n = 0$,
	if  $n_0 >  n_c$, one obtains $lim_{n \rightarrow \infty} n C_n =
	\infty $ and finally $lim_{n \rightarrow \infty} n C_n = 1$ if
	$n_0 = n_c$. We shall return to the interpretation of this
	critical value of $n_0$ later on. We shall see that the
	quantities $C_n$ correspond to the combinants of the multiplicity
	distribution and $n_0 = n_c$ critical value corresponds to
	the divergence of the mean boson multiplicity.

	Let us now proceed further with the solution of the model equations.
	The only remaining unsolved recurrence is eq.~(\ref{e:r.1}) which
	can be formally solved in a general manner, as follows.
 	Let us introduce the generating function of the $\omega_n$-s and
	as $G_{\omega}(z)$ and let us define an auxiliary $F(z)$ as follows:
 \bea
	G_{\om}(z) & = & \sum_{n=0}^{\infty} \omega_n z^n, \\
	F(z) & = & \sum_{n=0}^{\infty} (n+1) C_{n+1} z^n,
 \eea
	and observe that
 \bea
	F(z) G_{\om}(z) & = & \sum_{n = 0}^{\infty} \left[
			\sum_{k = 0}^n (k+1) C_{k+1} \omega_{n-k} \right]
			z^n, \\
	\int_0^z dt F(t)  G_{\om}(t) & = &
		\sum_{n=1}^{\infty} {\dst 1 \ov n} \left[
			\sum_{k = 0}^{n - 1}
			(k+1) C_{k+1} \omega_{n-k-1}\right]
			z^n = \sum_{n=1}^{\infty} \omega_n z^n,
 \eea
	and this last equation can be re-written   with the help of
	eq.~(\ref{e:r.1}) as
 \bea
	\int_0^z dt F(t)  G_{\om}(t)  & = & G_{\om}(z) - 1 , \\
	F(z) G_{\om}(z) & = & {\dst d \ov dz} G_{\om} (z) \\
	G_{\om}(z) & = &
		G_{\om}(0) \exp\left(\dst \int_0^z dt F(t) \right),
 \eea
	since the expansion coefficients of $F(z)$-s are known.
	The last step in the formal solution of eq.~(\ref{e:r.1})
	is the introduction of the generating function for the
	probability distribution $p_n$ as
 \bea
	G(z) & = & \sum_{n = 0}^{\infty} p_n z^n  \label{e:d.g}
 \eea
	which can be expressed with the generating function of the
	$\omega_n$ distribution with the help of eq.~(\ref{e:d.1}) as
 \bea
	G(z) & = & {\dst G_{\om}(z) \ov G_{\om}(1) }
 \eea
	which yields the formal solution for the generating function
	of the probability distribution of $p_n$ as follows:
 \bea
	G(z) & = & \exp\left( \sum_{n=1}^{\infty} C_n (z^n - 1) \right),
	\label{e:gsolu}
 \eea
	where the quantities $C_n$-s are the so called combinants of the
	probability distribution of $p_n$ and in our case their
	explicit form is known for any set of model parameters, as
	given by eqs. ~(\ref{e:c.s},\ref{e:gam.s},\ref{e:di.x},
	\ref{e:sigt},\ref{e:reff}). Note that this generating function
	depends on two variables, the parameter $n_0$ that controls the
	phase space density and the parameter $x =
	R_{eff}^2 \sigma_T^2$, eq.~(\ref{e:di.x}),
	that yields the available dimensionless phase-space volume.
	Alternatively, one may introduce another set of variables
	by defining $ t = n_0 / n_c$ and $ v = \sqrt{\gamma_{-} / \gamma_{+}}$,
	$ 0 \le v < 1$,
	to obtain a more transparent form of the probability generating
	function:
 \bea
	G(z;t,v) & = & \exp\left( \sum_{n=1}^{\infty} {\dst t^n\ov n}
		{\dst 1 \ov ( 1 - v^{n} )^3}  (z^n - 1) \right).
		\label{e:gsol}
 \eea
	We find that this multiplicity generating function
	is not corresponding to known discrete probability generating functions
	after inspecting a few standard textbooks on statistics
	that include a large number of probability generating
	functions, e.g. ref~\cite{johnson}.
	Thus we may assume that we have found a new type of
	probability generating function.
	We shall investigate its properties
	not only numerically but also analytically in certain simple limiting
	cases. Especially, we show that there is a critical value for
	$n_0$ at which a wave-packet version of Bose-Einstein condensation
	occurs that influences drastically the $p_n$ distribution too.
	We shall also show that this distribution  reduces to
	known type of distributions in certain limiting cases.

	The combinants were introduced to statistics in refs.~\cite{gyul-comb}
	and some of their properties were discussed recently in refs.
	~\cite{hegyi-c1}-\cite{hegyi-c4}. The auxiliary quantity
	$\omega_n$ is the probability ratio
 \bea
	\omega_n & = & {\dst p_n / p_0}.
 \eea
	(Note that in our model $p_0 = \exp( - \sum_n C_n) \ne 0$ ).
	The combinants up to a given order $n$ can be expressed
	as a combination~\cite{hegyi-c2} of the first $n$ probability ratios
	as
 \bea
	C_q & = & \omega_q -
		{\dst 1 \ov q} \sum_{n=1}^{q-1} n C_n \omega_{q-n},
 \eea
	hence their name. It is outside the scope of the present paper
	to study in detail the properties of the combinants and their
	relationships to other characteristics of the multiplicity
	distributions, like cumulants and factorial moments, and scaling laws
	of count probabilities. For a more detailed discussion on these
	general topics, we recommend refs.~\cite{hegyi-c1}-\cite{hegyi-c4}.

	Let us note, however, a general property of a multiplicity distribution
	that is given in terms of its combinants. Any probability generating
	function that is given in terms of its combinants that are non-negative
	can be re-written as a convoluted Poisson distribution,
	or compound Poisson distribution
	if $p_0 > 0$ as

 \bea
	G(z) = \exp( \overline{C} (H(z) - 1) ), \label{e:convpo}
 \eea
	where $ \overline{C} =  \sum_{n=1}^{\infty} C_n$ can be interpreted
	as the mean multiplicity of the Poisson-distributed clusters
	(or clans). For completeness, we note
	that the probability generating function for
	a simple Poisson distribution is
	$G_P(z) = \exp[ \, \overline{n} (z-1) ]$.
	 The particles within a single cluster are distributed
	according to the probability distribution
 \bea
	P_N = {\dst C_n \ov  \sum_{n=1}^{\infty} C_n}
 \eea
        and there is always at least one particle in any cluster i.e.
	$P_{N = 0 } = 0$.

	Another interesting property of this multiplicity distribution is that
	the generating function can be written as

 \bea
	G(z) = \prod_{n=1}^{\infty} \exp( C_n (z^n -1 ))
 \eea
	i.e. as an infinite product of generating functions of Poisson
	distributions of particle singlets, doublets, triplets etc, having a
	mean of $C_1$, $C_2$, $C_3$ etc respectively. The corresponding
	multiplicity distribution can therefore be expressed in terms of
	Poisson distributions of particle $n$-tuples with means of $C_N$
	in a multiple convoluted manner~\cite{gyul-comb,hegyi-c2}.
	This is a general property of any distribution that is given in
	terms of its combinants.

	Due to this property, the mean multiplicity can be expressed
	in terms of combinants as
\be
	\langle n \rangle = \sum_n n p_n = \sum_{i=1}^{\infty} i C_i
\ee
	We see that the finiteness of the mean multiplicity is related
	to the vanishing values of $i C_i$ for large $i$. This limit,
	on the other hand, can only be reached if $n_0 < n_c$.
	Thus one finds that $ \langle n \rangle < \infty $
	if $n_0 < n_c$ and $ \langle n \rangle = \infty $ if
	$ n_0 \ge n_c$. Thus $n_c$ can be interpreted as a critical
	value for the parameter $n_0$.
	We shall see that divergence of the mean multiplicity
	$ \langle n \rangle $ is related to condensation of the
	wave-packet modes with the highest multiplicities.

	In general, the factorial cumulant moments, $f_q$-s of the
	probability distribution can be expressed with the help of
	the combinants in a relatively simple and straightforward
	manner similarly to the results of refs.~\cite{hegyi-c3,bellac}:
\be
	f_q = {\dst d^q \ov dz^q} \ln G(z) |_{z=1} =
	\sum_{i=q}^{\infty} \, i \cdot (i-1) \cdot ...\cdot (i - q + 1) 
	\cdot C_i
\ee
	that are nothing else than the factorial moments within
	a single cluster multiplied with the average cluster
	multiplicity~\cite{hegyi-c3}.

	Note, however, that the multiplicity distribution we find can not
	only be considered as an infinite convolution of Poisson distribution
	of particle singlets, pairs triplets etc
	but can be re-written also as an infinite convolution of
	Bose-Einstein (or Negative Binomial) distributions with coupled
	parameters.

	In the exponent of $G(z)$ one may apply a negative binomial
	expansion of the terms
	\bea
		{\dst 1 \ov (\gamma_+^{(n/2)} - \gamma_-^{(n/2)})^3 }  &= &
		{\dst 1 \ov n_c^n} \sum_{k=0}^\infty
		{\dst (k+1) (k+2) \ov 2}
			\left( {\dst \gamma_- \ov\gamma_+}\right)^{n k / 2}
	\eea
	and with the help of this transformation, the generating function
	can be equivalently written as another infinite product:
 \bea
	G(z) & = &
		\prod_{k = 0}^{\infty}
		G_k(z)^{(k+1)(k+2) \over 2} , \label{e:gkbose}\\
	G_k(z) &= & {\dst 1 \ov
                \left[ 1 + \overline{n}_k (1 - z) \right]} ,\\
	\overline{n}_k & = & t (v^2)^k \, = \, \left({\dst n_0 \ov n_c} \right)
		\left({\dst \gamma_- \ov \gamma_+}\right)^k \, = \,
		2^{3/2} n_0
		{\dst (1 + x - \sqrt{ 1 + 2 x} )^k \ov
			(1 + x + \sqrt{ 1 + 2 x} )^{k + 3/2}}
 \eea
	This expression indicates that the probability distribution $p_n$
	can also be considered as an infinite convolution of generalized
	 Bose-Einstein distributions. The mean multiplicities of the
	single modes increase in a geometrical series.
	For any single mode of this convolution,
	the probability distribution can be written as
 \bea
	p_n^{(k)} & = & {\dst \overline{n}_k^n \over
			( \overline{n}_k + 1)^{(n+1)} },
 \eea
	and the mean multiplicity for that given mode can alternatively
	be written as
 \bea
	{\dst \overline{n}_k \ov \overline{n}_0} & = &
	\left({\dst \overline{n}_1 \ov \overline{n}_0} \right)^k
 \eea
	Thus the probability distribution can be considered {\it both}
	as a superposition of Poisson distributed independent
	clusters as well as an infinite convolution of
	Bose-Einstein distributions with coupled mean multiplicities.

	It is especially interesting to note, that in the
	dense Bose-gas limit, the $k=0$ term dominates
	eq.~(\ref{e:gkbose}), while in the rare Bose-gas limit
	each factor contributes with a similar weight from the infinite
	product in this equation. In contrast, one may consider the
	Poisson cluster decomposition as given by eqs.~(\ref{e:convpo},
	\ref{e:gsolu}) and in this type of decomposition,
	the first factor contributes to the rare gas limiting case but
	in the dense gas limiting case all factors are important.

	Let us make a few further remarks about the structure of the
	probability distribution $p_n$ in this class of models.
	Although this multiplicity distribution is already given in
	terms of its generating function by eq.~(\ref{e:gsol}),
	and in terms of its combinants, $C_n$ as given by eq.~(\ref{e:c.s}),
	a more explicit form for the multiplicities $p_n$ can also be
	given. In general, if the combinants are known, the probability
	ratios $\omega_n = p_n / p_0$ can always be expressed as
\be
	{\dst p_{n_1} \ov p_0} \,  = \,  \omega_{n_1}  \, =  \,
		\sum_{k = 1}^{n_1} {\dst 1 \ov k!}
		\left[
		\sum_{n_2 = k - 1}^{n_1 - 1} 
		\sum_{n_3 = k - 2}^{n_2-1} \, ... \,
		\sum_{n_k = 1}^{n_{k-1} } 
		C_{n - n_2} \, C_{n_2 - n_3} \, ...
		\, C_{n_{k-1} - n_k} \, C_{n_k}
		\right]
\ee
	Although this equation is seemingly more complicated than
	the recurrence	given for $\omega_n$ in eq.~(\ref{e:r.1}),
	it is still useful since the explicit form for $C_n$ can be inserted
	from eq.~(\ref{e:c.s}) to find that
\bea
	\omega_{n_1} &  = & t^{n_1}
		\sum_{k = 1}^{n_1} {\dst 1 \ov k!}
		\left\{
		\sum_{n_2 = k - 1}^{n_1 - 1} 
		\sum_{n_3 = k - 2}^{n_2-1} \, ... \,
		\sum_{n_k = 1}^{n_{k-1} } 
		\left[ \prod_{j=1}^{k-1}
		\left( 1 - v^{n_{j} - n_{j+1}} \right) \right]^{-1/3}
		\left( 1 - v^{n_{k}} \right)^{-1/3}
		\right\},
\eea
	where one can see that the $t = n_0 / n_c$ dependence of $ \omega_n
	\propto p_n$ is simple. This can be utilized to simplify the
	general evaluation of the $\omega_n$ probability ratios as
\bea
	\omega_n & = & t^n F_n(v) \\
	F_0(v) & = & 1 \\
	F_n(v) & = & {\dst 1 \ov n} \sum_{j=1}^n {\dst 1 \ov
		\left( 1 - v^j \right)^3 } F_{n-l}(v) .
\eea
	This is the simplest form that we can find for probability distribution
	in the general case (the probabilities are given by eq.~(\ref{e:d.1}) ).

	We investigate the probability  generating function $G(z)$
	as well as the multiplicities and the particle spectra
	 in more details in the  rare and in the dense gas limiting cases
	in the next sub-sections.

\subsection{Analytic Results for Rare Bose Gas \label{ss:rare}}
 In this section we consider large
 source sizes or large effective temperatures, i.e. we examine
 the combinants and the generating function of the probability
 distribution in the $ x >> 1 $ limiting case.
 Since the multiplicity distribution was obtained analytically in terms of the
 combinants, let us evaluate these to leading order in $ 1/x << 1$.
 One obtains that
 \bea
	\gamma_+ & \simeq &
		\left( {\dst x \ov 2} \right)^n
		+ n \left( {\dst x \ov 2} \right)^{ n  - {1\over 2}}, \\
	\gamma_- & \simeq &
		\left( {\dst x \ov 2} \right)^n
		- n \left( {\dst x \ov 2} \right)^{ n  - {1\over 2}}, \\
	\gamma_+^{ n} + \gamma_-^{ n} & \simeq &
		2 \left( {\dst x \ov 2} \right)^n, \\
	\gamma_+^{ n} - \gamma_-^{ n} & \simeq &
	    2 \, n \left( {\dst x \ov 2} \right)^{n - {1\over 2}}.
 \eea
	These relationships can be substituted to the solution of the
	recurrence relations, eqs.~(\ref{e:h.s}-\ref{e:g.s})
	to obtain that
 \bea
	h_n & = & {\dst 1 \ov (\pi \sigma_T^2)^{(3/2)} } {\dst 1 \ov n^{3/2} }
	\left( {\dst x \ov 2 } \right)^{{3\over 2}(1-n)}, \\
	g_n & = & {\dst x \ov n \sigma_T^2 } = {\dst R_{eff}^2 \ov n},\\
	C_n & = & { \dst n_0^n  \ov n^4 }
		\left({\dst 2 \ov x} \right)^{{3\over 2}(1-n)}.
	\label{e:c.rare}
 \eea
	Note that the determination of $a_n$ is a bit tricky, since  a
	finite sub-leading part remains that is difficult to determine as
	the leading orders cancel.
	However, there is an important constraint between
	$C_n$ and $a_n$ as given by eq.~(\ref{e:a.2}).
	This constraint is satisfied by
 \be
	a_n  = {\dst n \ov 2 \sigma_T^2 } + {\dst R_{eff}^2 \ov 2 n}.
 \ee
	Inserting these to the definition~(\ref{e:a.1}) one obtains that
 \bea
	G_n(\bk_1,\bk_2) & = &
		{\dst n_0^n \ov (n \pi \sigma_T^2)^{3/2} }
		\left({\dst 2 \ov x}\right)^{{3\over 2}(n-1)}
		\exp\left[
		 - {\dst n \ov 2 \sigma_T^2 }( \bk_1^2 + \bk_2^2)
		- {R_{eff}^2 \ov 2 n}
			\left(\bk_1 - \bk_2\right)^2 \right].
	\label{e:g.rare}
 \eea
	Both the inclusive and the exclusive momentum distributions
	will be built up from the auxiliary quantities
	$G_n(\bk_1,\bk_2)$, and the leading order
	behavior is given by $G_1(\bk_1,\bk_2)$
	in this rare gas limiting case, with first order corrections from
	$G_2(\bk_1,\bk_2)$.

	Eq. ~(\ref{e:c.rare})
	indicates, that the leading order result for the combinants
 	in the $ x >> 1 $ limiting case is
 \bea
	C_n & = & \delta_{1,n} n_0
 \eea
	and the first sub-leading correction is given by
 \bea
	C_n & = & \delta_{1,n} n_0 + \delta_{2,n} \,
		{\dst n_0^2 \ov 2 (2 x)^{3/2} }
 \eea
	Thus, to leading order, the probability distribution can be considered
 	as a Poisson distribution of singlets with a sub-leading correction
 	that yields a convolution of Poisson-distributed doublets.
	Similarly to this, the Poisson-doublets will modify the single-particle
	momentum distribution as well as the two-particle correlation function.
	We consider these modifications in two manner: {\it i)} by
	evaluating the leading order corrections only and {\it ii)}
	by summing up certain sub-leading corrections.

\subsubsection{Very rare gas}

	In the rare gas limiting case, the leading order Poisson distribution
	can be utilized to solve the model completely as follows:
 \bea
	\omega_n & = & {\dst n_0^n \ov n!}, \\
	p_n & = & {\dst n_0^n \ov n!} \exp( - n_0), \\
	G_n(\bp,\bq) & = & \delta_{1,n} \,
		{\dst n_0 \ov ( \pi \sigma_T^2 )^{(3/2)} } \,
		\exp\left( - {\dst R_{eff}^2 \ov 2} (\bp - \bq)^2
		- {\dst \bp^2 + \bq^2 \ov 2 m T_{eff} } \right), \\
	P^{(n)}_1({\bk}) & = & P^{(1)}_1({\bk}) \, = \,
		{\dst 1 \ov ( 2 \pi m T_{eff} )^{(3/2)} } \,
		\exp\left( -{\dst {\bk}^2 \ov 2  m T_{eff} } \right), \\
	P_1({\bk}) & = & P^{(1)}_1({\bk}) \, = \,
		{\dst 1  \ov ( 2 \pi m T_{eff} )^{(3/2)} } \,
		\exp\left( -{\dst {\bk}^2 \ov 2  m T_{eff} } \right), \\
	P_2(\bk_1,\bk_2) & = &
			P_1^{(1)}(\bk_1)\,  P_1^{(1)}(\bk_2) \,
			\left[ 1 +
		\exp\left( - R_{eff}^2 (\bk_1 - \bk_2)^2 \right)\right], \\
	C_2(\bk_1,\bk_2) & = &   1 +
			\exp\left( - R_{eff}^2 (\bk_1 - \bk_2)^2 \right)
			 . \\
 \eea
	Thus the very rare gas limiting case is very simple and solvable
	completely in an analytical manner. The multiplicity distribution
	is a Poisson distribution, the mean multiplicity coinciding with
	$n_0$. The stimulated emission does not influence the probability
	distribution for the ``unsymmetrized" $p_{0,n}$ that we assumed to
	have a Poisson form. There is only a normalization factor that
	appears as a modification
	to the single-particle momentum distributions, and, apart from
	an overall normalization factor ${\cal N}_c$ the two-particle
	inclusive correlation function takes a Gaussian form with an
	effective radius parameter $R_{eff}$. For large mean multiplicities,
	the overall normalization factor approaches unity since
          $ \lim_{n_0 \rightarrow \infty} {\cal N}_c = 1$.

	To our best knowledge, no similar evaluation of the
	two-particle inclusive correlations from the recurrence relations of
	S. Pratt was performed before, only
	exclusive correlation functions	  were evaluated numerically.
	Thus, this is the first time as far as we know that the exclusive
	momentum distributions are summed up and the inclusive particle
	spectra and the inclusive correlations are determined
	from the plane-wave model, too. Our method seems to be powerful enough
	to do certain analytical calculations even for systems where the
	numerical evaluation of the inclusive distributions were too
	tedious to perform earlier. Let us also stress that
	these results correspond to a multi-particle wave-packet system.

	One may argue that the solution presented above is rather trivial
	and that not much is learned about the nature of the symmetrization
	effects here since they cancel from the leading order results in the
	rare Bose-gas limiting case. But the above simple results
	became very useful when one determines the next to leading order
	results in the rare gas limit, and it is necessary to know what
	is the reference point and what kind of changes happen if the
	leading order contribution from the symmetrization is taken into
	account.

\subsubsection{The rare gas limiting case}
	Let us evaluate now the leading order corrections to the
	Poisson limiting case.
	This corresponds to keeping the first order corrections in $1/x$.
	The probability generating function reads as
\be
	G(z)  =  \exp\left( n_0 (z - 1) + C_2 (z^2 - 1) \right) \,
		\simeq \,
		\exp\left( n_0 (z - 1) \right) \,
		\left( 1 + {\dst n_0^2 \ov 2 (2 x)^{(3/2)}
		}(z^2 - 1) \right)
\ee
	which yields the following multiplicity distribution:
\be
	p_n = {\dst n_0^n \ov n!} \exp(-n_0) \,
		\left[ 1 + {\dst n(n-1) - n_0^2 \ov 2 (2 x)^{(3/2)}
			} \right].
	\label{e:pn.sol}
\ee
	We see that the leading order Poisson multiplicity distribution
	is modified in such a manner that the corrected
	 multiplicity distribution is depleted in the
	small $n$ region, where $ n (n-1) < n_0^2$ and enhanced in the
	high multiplicity region, where $ n (n-1) > n_0^2 $,
	since
\be
	{\dst p_n \ov p_{0,n} } = 1 + {\dst n(n-1) - n_0^2
			\ov 2 ( 2 x)^{(3/2)} }.
\ee
	 This {\it shift to large multiplicities } is more enhanced for
	smaller values of $x$ i.e. for higher phase-space density.
	The mean multiplicity can be calculated as
\be
	\langle n \rangle = {\dst d\ov dz} G(z)|_{z=1}
	= n_0  + {\dst n_0^2 \ov (2 x)^{(3/2)}}
	\label{e:n.ave.rare}
\ee
	which increases in this leading order  calculation
	 quadratically with increasing values of
	$n_0$ due to the influence of Poisson-doublets.
	Note also the strong rise of the mean multiplicity with
	decreasing values of $x$ i.e. with increasing phase-space
	density of the wave-packets.

	Evaluation of similar leading order corrections to the spectra
	and the correlations is straight-forward.

	The $n$-particle exclusive and inclusive number-distributions have
	simpler form than the corresponding probabilities, hence we
	indicate here these quantities:
\bea
	N_1^{(n)}(\bk) & = & {\dst n \ov ( \pi \sigma_T^2)^{(3/2)} }
			\exp\left( -{\dst  \bk^2  \ov \sigma_T^2} \right) + \\
	\null & \null &
			{\dst n(n-1) \ov (2 x)^{(3/2)} }
		\left\{ \left[ {\dst 2 \ov \pi \sigma_T} \right]^{(3/2)}
		\, \exp\left( - {\dst 2 \ov \sigma_t^2} \bk^2 \right)
		- \left[ {\dst 1 \ov \pi \sigma_T^2} \right]^{(3/2)}
		\exp\left( - {\dst 1 \ov \sigma_T^2} \bk^2 \right)
		\right\} \label{e:n.x.rare} \\
	N_1(\bk) & = & {\dst n_0 \ov ( \pi \sigma_T^2)^{(3/2)} }
		\exp\left( -{\dst  \bk^2  \ov \sigma_T^2} \right) +
		{\dst n_0^2 \ov (\pi \sigma_T^2 x )^{(3/2)} }
		\, \exp\left( - {\dst 2 \ov \sigma_t^2} \bk^2 \right)
		\label{e:n.i.rare}
\eea
	Note that these quantities are properly normalized
	according to eqs.~(\ref{e:norm1}), where the
	mean multiplicity is given by	eq.~(\ref{e:n.ave.rare}).

	The interpretation of these results is straight-forward:
	{\it i) The single-particle inclusive momentum distribution
	is enhanced at low momentum.}

	The enhancement is proportional
	to ${\dst n_0^2 \ov x^{3/2} } =
	{\dst n_0^2 \ov (2 m T_{eff} R_{eff}^2)^{(3/2)} } $ which is
	the mean number density of pairs in the phase-space volume
	that is available for a single particle in the rare gas limit.

	{\it ii) The single-particle exclusive momentum distribution
	of rank $n$  is enhanced at low momentum,
	the enhancement being proportional to the number of pairs,
	$n(n-1)$. } This predicts a characteristic dependence in
	the exclusive momentum distribution as a function of $n$,
	that is due to the Bose-Einstein symmetrization.
	Such dependences could be studied in the event by
	event analysis to be performed by the NA49 Collaboration, for
	example.

	The evaluation of the Bose-Einstein correlation functions
	in this limiting case is rather involved and will be published
	separately~\cite{cs-z}.

	Let us turn our attention to the opposite limiting case,
	the ultra-dense Bose-gas of wave-packets.

\subsection{Analytic Results for Highly Condensed Bose Gas \label{ss:cond}}

	This limiting case corresponds either to a
	small effective radius parameter
	or a low effective temperature of the multi-boson wave-packet
	system. Formally, it corresponds to the $ x = 2 m T_{eff} R_{eff}^2
	<< 1 $ limiting case. In this case, one can determine the
	leading order expression for the combinants of the probability
	distribution as
\bea
	\gamma_+ &  = & 1 + x + {\cal O}(x^2), \\
	\gamma_- &  = & 0 + {\cal O}(x^2), \\
	n_c & = &  n_c(x) \, \simeq \,
			(1 + x )^{3/2} \simeq 1 + {\dst 3 \ov 2} x \\
	C_n & = &  {\dst 1 \ov n} \, \left({\dst n_0 \ov n_c}\right)^n
		\simeq
		{\dst n_0^n \ov n}  - {\dst 3 x \ov 2} n_0^n + {\cal O}(x^2).
\eea
	Thus the probability generating function can be written in the
	condensed Bose-gas limiting case as
\bea
	G(z) & \simeq & {\dst 1 \ov ( 1 - \overline{n} (1 - z)) } \\
	\overline{n} & = & {\dst n_0 \ov n_c(x) - n_0 }
\eea
	which corresponds to a Bose-Einstein distribution in the
	$x << 1 $ limiting case with a mean multiplicity that diverges
	if $n_0 $ reaches the value of $n_c \simeq (1 + 3x/2)$.
	In this $ x << 1 $ limiting case,
	the multiplicity distribution shall be given by a Poisson distribution,
	and the model can be completely solved analytically as follows:
\bea
	p_n & = & {\dst \overline{n}^n \ov ( \overline{n} + 1)^{n+1} }, \\
	\omega_n & = & \left( {\dst n_0 \ov n_c}\right)^n  \, =
		\, {\dst  \overline{n}^n \ov ( \overline{n} + 1)^n }, \\
	G_n({\bp},{\bq}) & = &
		{\dst \overline{n}^n \ov ( \overline{n} + 1)^{n+1} }
		G_1({\bf p},{\bf q}) \, = \,
		{\dst \overline{n}^n \ov ( \overline{n} + 1)^{n+1} }
		 \, {\dst 1 \ov (\pi \sigma_T^2)^{(3/2)} } \,
		\exp\left( - {\dst {\bf p}^2 + {\bf q}^2  \ov 2 \sigma_T^2}
		\right), \\
	P^{(n)}_1({\bk}) & = & P^{(1)}_1({\bk})
		 \, = \,
		{\dst 1 \ov ( 2 \pi m T_{eff} )^{(3/2)} } \,
		\exp\left( -{\dst {\bk}^2 \ov 2  m T_{eff} } \right), \\
	P_1({\bk}) & = & P^{(n)}_1({\bk})
		 \, = \,
		{\dst 1 \ov ( 2 \pi m T_{eff} )^{(3/2)} } \,
		\exp\left( -{\dst {\bk}^2 \ov 2  m T_{eff} } \right), \\
	P_2(\bk_1,\bk_2) & = & P_1(\bk_1) \, P_1(\bk_2), \\
	C_2(\bk_1,\bk_2) & = & 1.
 \eea
	Note that this solution is obtained by summing up leading order
	contributions in the small $x$ region.
 In this limiting case, corresponding to $x = 2 m R_{eff} T_{eff} << 1$,
 essentially all the wave-packets remain condensed in the same wave-packet
 state as reflected by the shape of the single-particle inclusive spectrum
 and by the vanishing enhancement of the inclusive Bose-Einstein correlations.
 The mean multiplicity diverges as $n_0$ approaches $n_c \simeq
	1 + {\dst 3 x \ov 2}$. This corresponds to a very small wave-packet
  system where stimulated emission can dominate the particle emission
 even for very small values of the parameter $n_0$.
 The multiplicity distribution is a Bose-Einstein distribution,
 the parameter $n_0$ can be interpreted as  a measure of $  p_n / p_{n+1} < 1$
	in units of $n_c$,
 the inclusive and exclusive single-particle distributions coincide
 and there is no enhancement in the two-particle inclusive correlation
	function. Due to this reason, the solution corresponds to a
 completely coherent behavior, when all the particles are emitted in the
 same wave-packet state. It is interesting to observe, that the coherent
 behavior in this picture (defined by the vanishing enhancement in
	the two-particle inclusive correlation function) appears simultaneously
 with a thermalized Bose-Einstein distribution of the multiplicity
 distribution for $\bk_1 \ne \bk_2$. Thus the effective coherence appears
 due to the negligibly small effective source-size $R_{eff} \rightarrow 0$,
 or a very small effective temperature $T_{eff} \rightarrow 0$.
 In these limiting cases,
the model corresponds to a thermalized source prepared in such a manner
 that all the wave-packets are created in the same state and their overlap
 is maximal. This is the reason why thermal behavior and
 coherence can appear simultaneously in this picture.
 It is interesting to note that the single-particle inclusive
 and exclusive momentum distributions coincide and are described by
 a Boltzmann distribution with an effective temperature $T_{eff} = T +
{	\sigma^2 \ov 2 m}$ that picks up a contribution from the
 temperature of the source and from the momentum-width of the wave-packets.
 This shape of the momentum distribution is forced by the Gaussian
ansatz for the momentum-space representation of the wave-packets and
the Gaussian distribution of the mean momentum of the packets.

 In this and the previous sub-sections,
  analytical solutions
 to the multi-particle wave-packet system are presented. 
 These analytic solutions yield
 new result not only in the study   of multi-particle wave-packet problems,
 but also of the plane-wave problem that corresponds to the
 wave-packet problem, obtainable 
 by the substitution $R_{eff} \rightarrow R$ 
 and $T_{eff} \rightarrow T$.

\subsection{
  Other Results for the Critical Multiplicity
	\label{ss:onset}}
 The critical behavior, related to the divergence of the 
 mean multiplicity in the original plane-wave system
 was discussed first in   ref.~\cite{plaser}.
 In this pioneering work,
 S. Pratt published a formula for the onset of the condensation,
 determining $n_c$ the critical number of the unsymmetrized pions
 from the inspection of certain ring-diagram. 
 Although Figure 1 of ref.~\cite{plaser} was evaluated with the
 correct expression~\cite{ppcom},
 \be
	\eta_c \, = \, \left[ {\dst 1 \ov 2} +
	\Delta^2 R^2 +
	\sqrt{ \Delta^2 R^2 + {\dst 1 \ov 4} } \right]^{(3/2)},
	\label{e:ncpratt}
 \ee
 this expression appeared in the paper in
 a misprinted manner, eq. (9) of ref.~\cite{plaser}, 
 where $\eta_c = n_c$ and $\Delta = m T$.
 Note that the steps to derive eq. (9) were not given in
 ref.~\cite{plaser} and the misprinted formula depends also on
  $p_0 = {\bp}^2/(2 m)$. 
 Our result on the critical density for the condensation we define
 can be written as
\be
	n_c = \left[ {\dst 1 + 2 m T_{eff} R_{eff}^2 + \sqrt{
	1 + 4 m T_{eff} R_{eff}^2} \over 2} \right]^{(3/2)},
	\label{e:nccstjz}
\ee
 which corresponds to the correct formula of eq.~(\ref{e:ncpratt})
 when going to the wave-packet system from the plane-wave system
 with the help of the replacements
 $\eta_c \rightarrow n_c$, $ \Delta = m T \rightarrow m T_{eff}$
 and $ R \rightarrow R_{eff} $.

\begin{figure}
          \begin{center}
          \leavevmode\epsfysize=3.0in
          \epsfbox{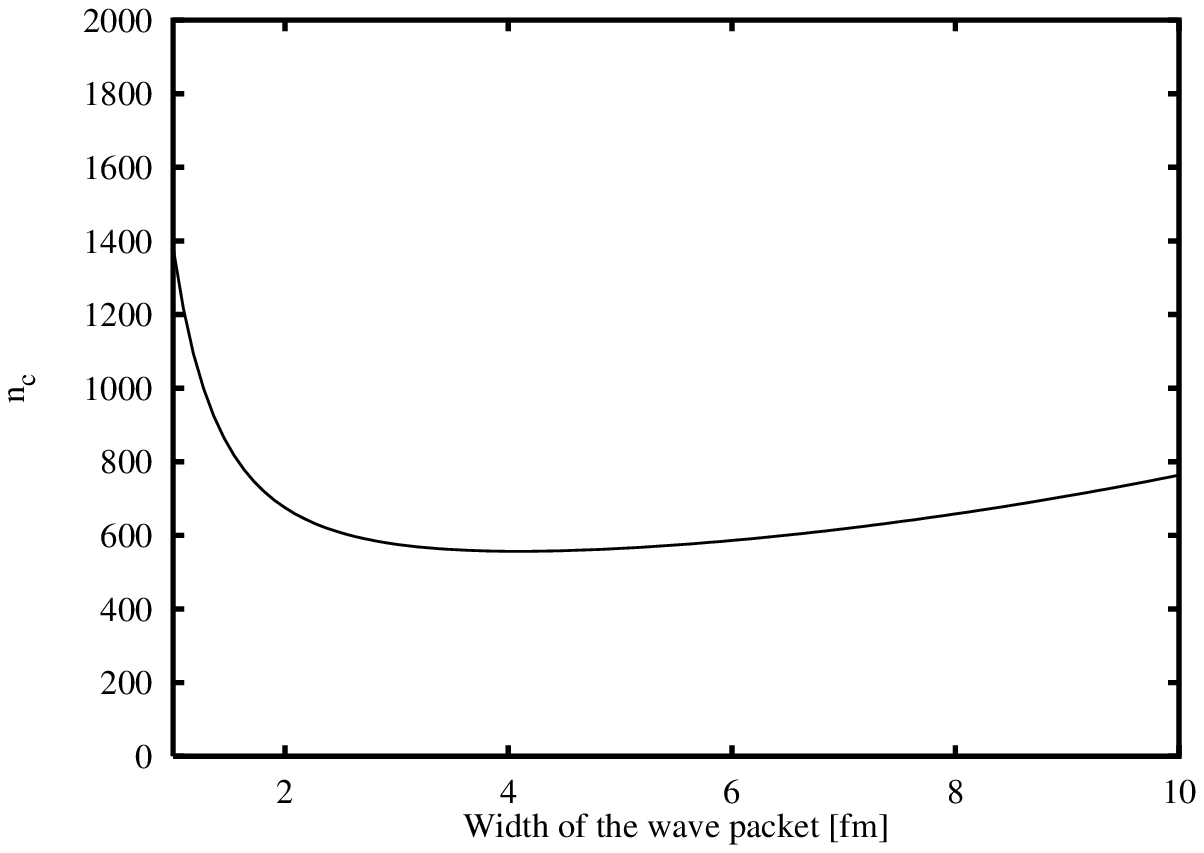}
          \end{center}
 \caption{  
}
\label{f:1}
\end{figure}

On Figure~(\ref{f:1}) we indicate the critical pion multiplicity
as a function of wave-packet size. 

 It is important to discuss how the critical value
 of the parameter $n_0$ can be determined.
 We were interested in finding that limiting case, when the
 explosion of the  (symmetrized) $p_n$ probabilities just happens.
 This we call the Bose-Einstein condensation point and we define
 it physically as the curve in the $n, R_{eff}, T_{eff}$ parameter space
 along which the decrease of the (unsymmetrized ) $p_n^{(0)}$ probabilities
 is just compensated by the stimulated emission of bosons.
 With other words, this is the point where a strong stimulated emission
 is build-up, which corresponds to the onset of the lasing mode.

 Thus the condensation point is defined as
\bea
	n p_n (n_c) &= & (n+1) p_{n+1}( n_c)
\eea
	which is to be satisfied in the large $n$ limit only.
 It is clear from this definition, that when this condition is satisfied,
 the mean multiplicity has to diverge.

	Let us first determine the critical boson multiplicity
	for the onset of the Bose-Einstein condensation of the
	wave-packets into the same wave-packet state.
 From $n p_n = (n+1) p_{n+1}$ in the large $n$ limit it follows that
 $n \omega_n = (n+1) \omega_{n+1}$.  From this it follows that $n C_n =
 (n+1) C_n$ in the large $n$ limiting case.
 Comparing this to eq.~(\ref{e:gam.s}), one finds that
at the onset of Bose-Einstein condensation, as noted before,
happens at
\bea
	n_0 & = & n_c \, = \,
		\left[{\dst 1 + x + \sqrt{1 + 2 x} \ov 2} \right]^{(3/2)}
\eea
 where $ x = R_{eff}^2 \sigma_T^2  =  2 m T_{eff} R_{eff}^2$ 
 is a dimensionless measure of the
 size of phase-space cell occupied on average by a single quanta.
 This expression is identical to eq.~(\ref{e:nccstjz}).
 One may determine this critical $n_c$ parameter of the distribution
 as a function of $\{R_{eff}, T_{eff}\}$ or as a function of
 $\{R, T, \sigma\}$.

Thus the parameter $n_0$ can be compared to its critical value
 $n_c$ and if $n_0 < n_c$, $\lim_{n \rightarrow \infty} p_n = 0$
but above the critical multiplicity
$n_0 >  n_c$, $\lim_{n \rightarrow \infty} p_n = \infty$
i.e. stimulated emission over-compensates the decrease of the
unsymmetrized emission probabilities and a coherent, laser-like
behavior occurs.

 Thus the ratio $n_0/n_c$ controls the competition between the
	stimulated emission and the decrease of the production
	probabilities of the unsymmetrized emission, and $n_c$
can be interpreted as a critical pion density at which the condensation
of wave-packets starts to appear.

\section{Numerical Results for Multi-Particle Wave-Packets
	\label{s:numerics}}

In this section we present some numerical results about the effects
of multi-particle symmetrizations on the multiplicity distribution,
on the single-particle exclusive momentum distribution
and  on the two-particle exclusive correlation functions.
Although these results are more readily interpretable when one
can rely on the analytical insights gained in the previous section,
a qualitative understanding of the multi-particle symmetrization effects
on these observables can be obtained from a direct inspection of these
figures.
Let us investigate first the modification of the multiplicity distribution,
$p_n$, due to the overlap of the wave-packets.

\begin{figure}
          \begin{center}
          \leavevmode\epsfysize=3.0in
          \epsfbox{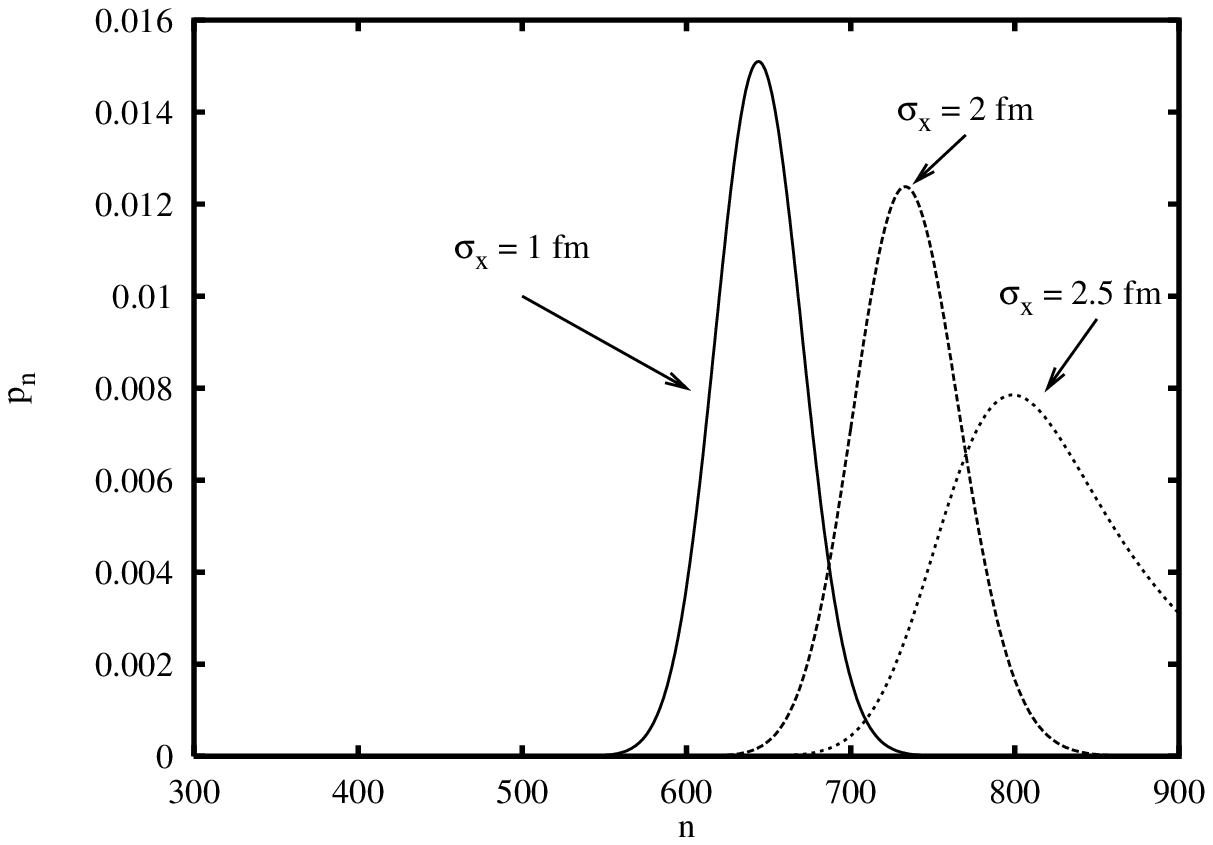}
          \end{center}
 \caption{  
}
 \label{f:2}
\end{figure}

\begin{figure}
\null
          \vspace{-0.4 truecm}
          \begin{center}
          \leavevmode\epsfysize=3.0in
          \epsfbox{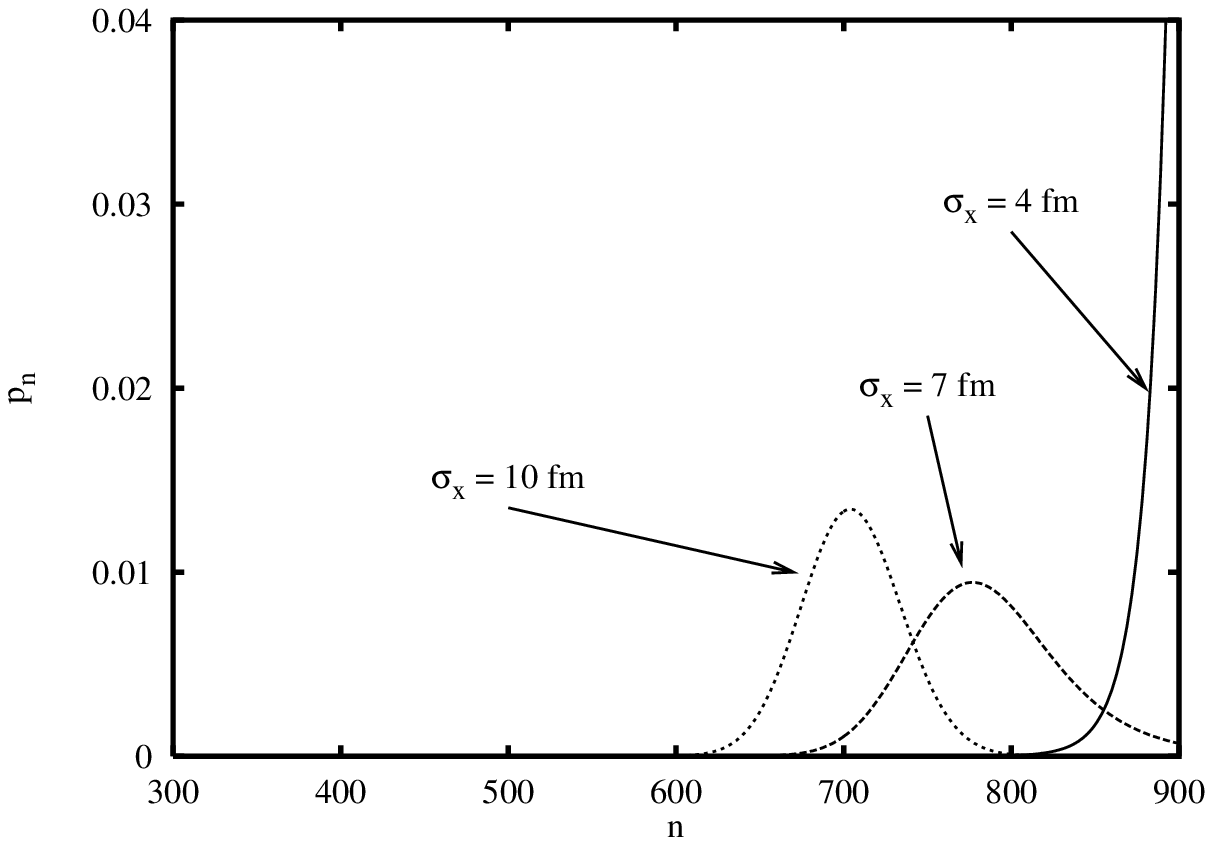}
          \end{center}
 \caption{  
}
 \label{f:3}
\end{figure}

In Figs.~\ref{f:2} and \ref{f:3} the multiplicity distribution,
   eq.(60) is plotted for various wave packet sizes.
The ``multiplicity parameter",
 $n_0 $, was fixed to $n_0 = 600 $. These Figures clearly demonstrate,
that the distribution function approaches a divergent region
for wave packet sizes
in the vicinity of $ \sigma_x = 4 fm $ for this set of parameters.
This is due to the fact that the overlap between the wave-packets is
controlled by the wave-packet size and if the overlap reaches a critical
magnitude, stimulated emission of wave-packets in similar wave-packet
states leads to a multiplicity distribution that diverges for large values of
$n$. As the size of the overlap of the multi-particle wave-packets
is increased, the multiplicity distribution becomes more and more
different from the Poissonian limit, and it is shifted to larger
multiplicities in accordance with our leading order analytical result given by
eq.~(\ref{e:pn.sol}).

\begin{figure}
          \vspace{-0.4 truecm}
          \begin{center}
          \leavevmode\epsfysize=3.0in
          \epsfbox{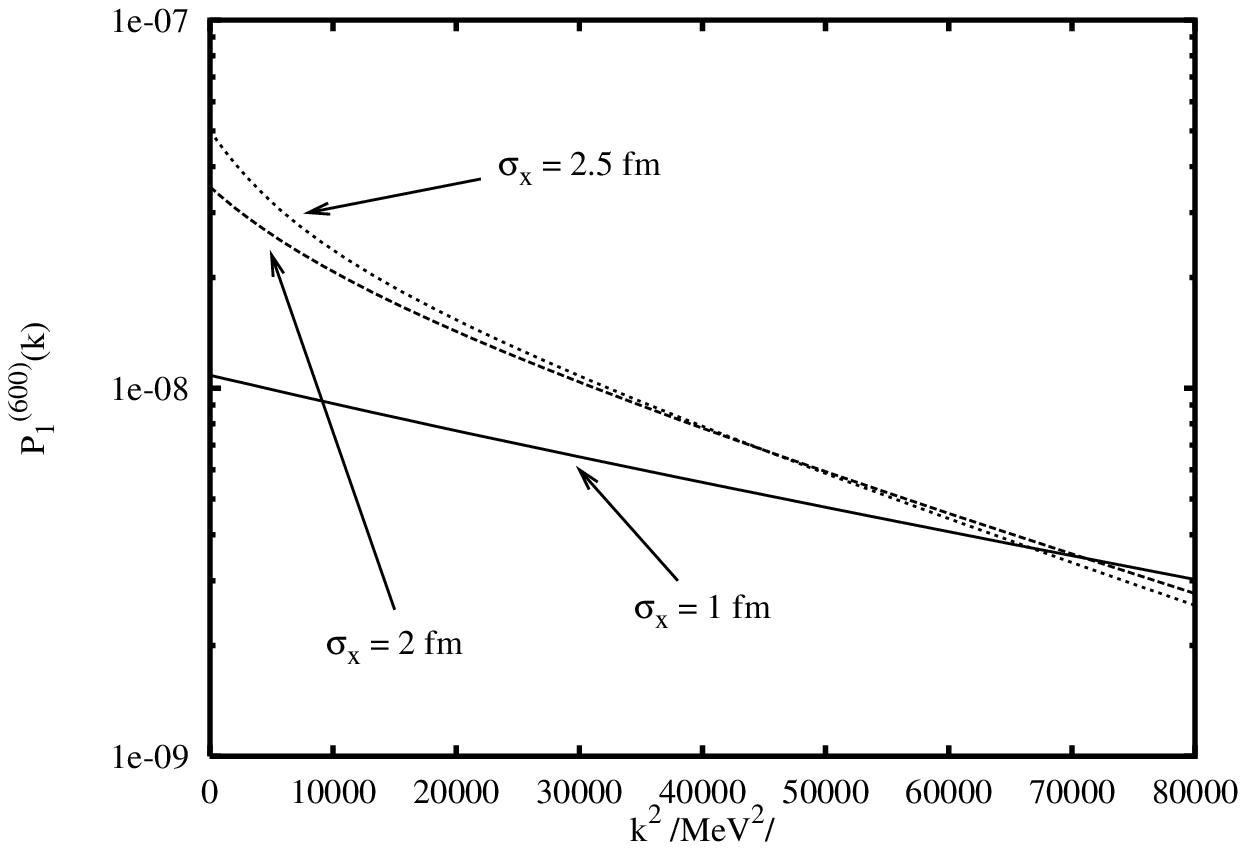}
          \end{center}
          \vspace{-0.3 truecm}
 \caption{ 
}
	\label{f:4}
\end{figure}

\begin{figure}
\null
          \vspace{-0.4 truecm}
          \begin{center}
          \leavevmode\epsfysize=3.0in
          \epsfbox{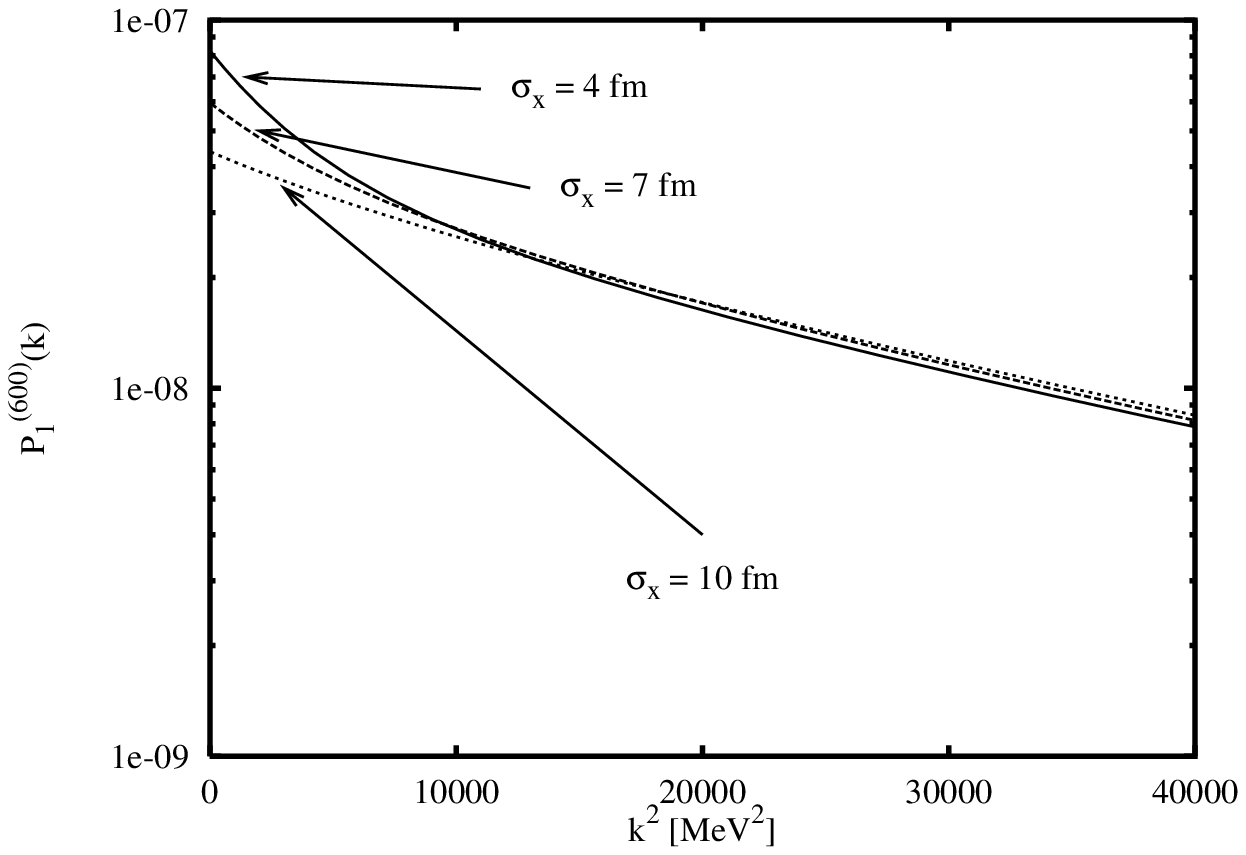}
          \end{center}
          \vspace{-0.3 truecm}
 \caption{ 
}
	\label{f:5}
\end{figure}

Investigating the single-particle exclusive momentum distributions,
our analytical results about the development of a low-momentum enhancement
is confirmed also numerically, as can be seen on
 Figures~\ref{f:4} and \ref{f:5}. It is also worth mentioning that
the modification of the slope parameter of the high momentum part of
the distribution is significant only when the spatial width of the
wave-packets is small. This happens since the effective slope parameter,
$T_{eff}$ is given by eq.~(\ref{e:teff}) and the contribution of the
wave-packets to this quantity is large only if the momentum-width of
the packets is large i.e. their spatial width is small.

\begin{figure}
          \begin{center}
          \leavevmode\epsfysize=3.0in
          \epsfbox{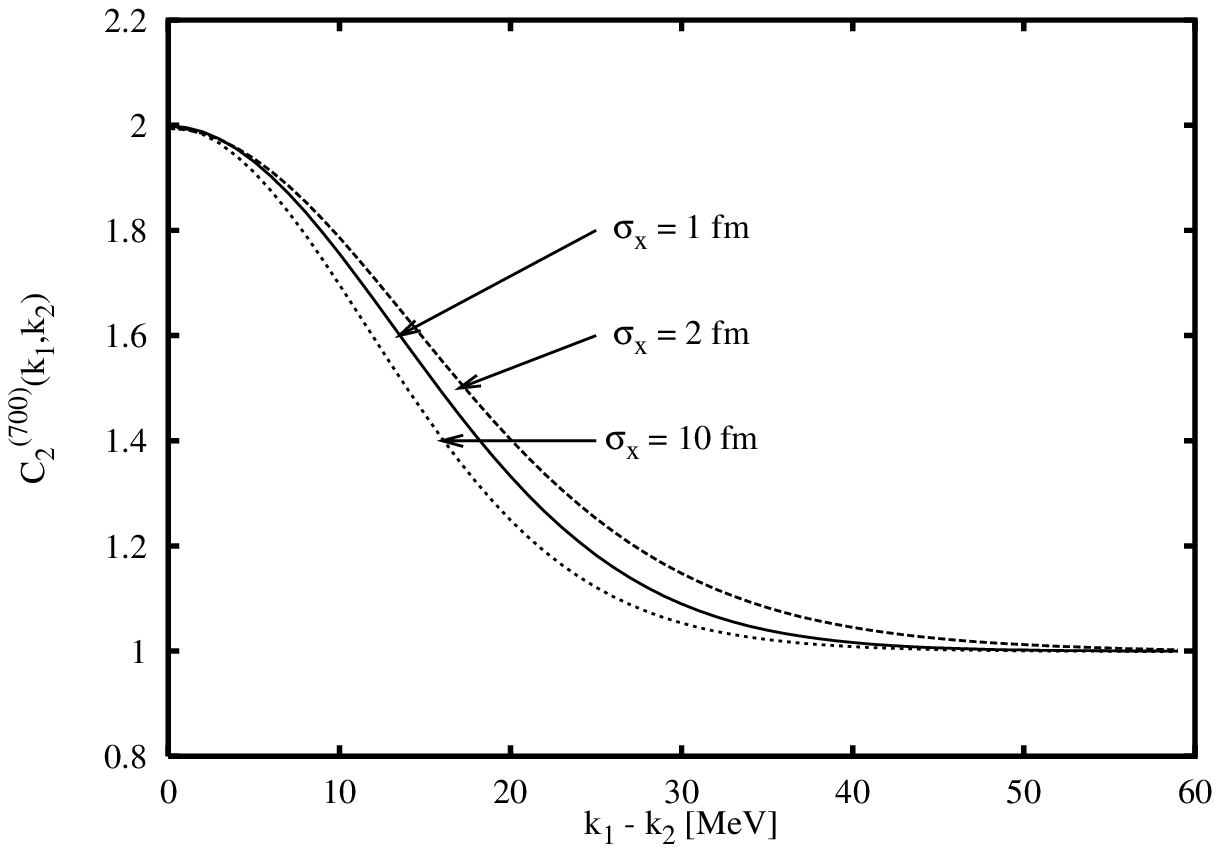}
          \end{center}
 \caption{ 
}
	\label{f:6}
\end{figure}

\begin{figure}
          \begin{center}
          \leavevmode\epsfysize=3.0in
          \epsfbox{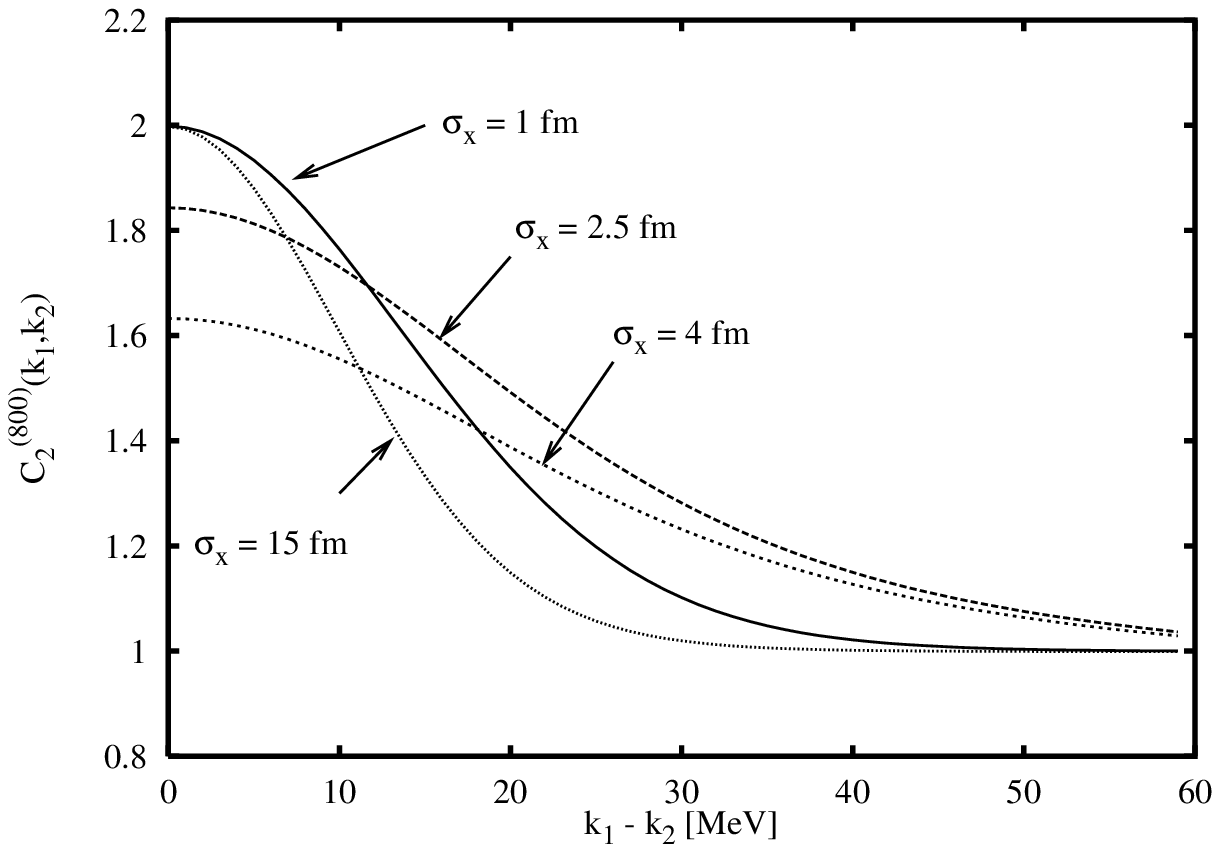}
          \end{center}
 \caption{ 
}
	\label{f:7}
\end{figure}

The effect of the wave packet size on the two-particle exclusive
 correlation  functions is shown on Figs.~\ref{f:6} and \ref{f:7}.

The virtual size of the source, $R_v$, is assumed to
 be $R_v=1/|\Delta \bk |_{1/2}$, where $| \Delta \bk |_{1/2}$ is the momentum
difference, at which the correlation  $C(\Delta \bk ) -1 $ drops to half of
its intercept value, $ C(0) -1 $ .Our first expectation would be, that $R_v$ is
a monotone increasing function of $\sigma_x$. However, as Fig.~\ref{f:6} shows,
this expectation is not fulfilled.
This effect is even more pronounced in Fig.~\ref{f:7}.
The explanation of this behavior
can be given as follows. From Figs.3 and 4. we have seen, that the
system approaches the formation of a  condensate at $\sigma_x=4.0  $ fm.
At this size, the overlap between the wave packets within
the "virtual volume" is maximal.
This case corresponds to the smallest ``virtual" radius
parameter of the exclusive correlation function and to the smallest
value of its intercept parameter $\lambda = C(\bk,\bk) - 1$.

It is also clear, that the condensation effects the low momentum pions.
Thus as we changed the mean pion momentum from 100 MeV to 50 MeV,
we included more and more from the condensed part, which is responsible
for the strange behavior.

	Note that the ``virtual" radius parameter first decreases than
increases with increasing values of the wave-packet size $\sigma_x$.
Similarly, the intercept parameter $\lambda = C_2(\bk,\bk)$
first decreases then it  increases back to its conventional value of 1,
for increasing size of the wave-packets.
As we have seen in the ultra-dense limiting case, $\lambda = 0$
is also possible for high densities.

\section{Summary, Conclusions and Outlook}

In this paper a consequent quantum mechanical description
of multi-boson systems is presented, using properly normalized
projector operators to overlapping multi-particle wave-packet states for
bosons.   Hundreds of overlapping wave packets lead to a difficult problem.
We found, however, a possible configuration, for which an exact
solution is obtained. In our description no phase averaging is used.
The mathematical problem imposed by the large number of source points
is dealt with an algebraic procedure, similar to that invented
by S.Pratt in ref.~\cite{plaser}.
The  effects arising from the multi-particle
symmetrization and from the finite width of the wave packet, $\sigma_x$,
are shown for a system that has a radius $R$ and
a freeze-out temperature $T$ similar to a fireball that could
be formed  at mid-rapidity in Pb + Pb collisions at CERN  SPS.
The effects that depend on the wave-packet size are as follows:

{\it 1).} The critical
density of pions, at which the condensation appears,
 is a function of $\sigma_x$ , and one finds 
that a pion-laser is formed at around $\sigma_x = 4 $ fm, 
when assuming $n_0 = 600$, $R = 11$ fm and $T = 120$ MeV.

{\it 2).} The multiplicity distribution shifts toward higher and higher
multiplicities  as $\sigma_x$ is increased towards $ 4 $ fm. In
a range of wave-packet widths that include 4 fm,
the multiplicity distribution explodes, i.e. we approach the
condensation point. As $\sigma_x$ is further increased, the
average multiplicity decreases back again to normal values.

{\it 3).}
The single particle momentum distribution shows a
very large inverse slope parameter for small $\sigma_x$, say
$\sigma_x = 1 $ fm. This is a trivial effect: in the Fourier
transform of a narrow wave packet there is a large contribution
from high momentum components, which shows up in the single
particle momentum distribution. Further, as $\sigma_x $
approaches the 4 fm value, the low momentum part in the spectrum
becomes more and more enhanced. This is understandable, if we keep in
mind, that the Bose condensation is approached first for the low
momentum part of the spectrum. As $\sigma_x$ is further
increased, the spectrum approaches again a Boltzmann
distribution. This can be understood, if we take into account,
that for larger $\sigma_x$ the virtual volume increases and
thus the density of pions within this volume decreases.

{\it 4)} The two particle correlation function shows an interesting
behavior. Two features are to be emphasized: {\it i)} The virtual
radius first decreases and later increases with increasing
$\sigma_x$. This effect is larger as the mean momentum of the
two pions decreases. {\it ii)} The intercept value, $
C(| \bk_1 - \bk_2 | = 0) $, decreases as $\sigma_x$ is increased
from 1 fm to 4 fm, and increases again as $\sigma_x $ is
further increased. These effects are caused by the increase
and decrease of pion density within the "virtual volume",
which is influenced by the
distance to onset of the divergence of the mean multiplicity, i.e.
the onset of the laser  (or condensation) mode.

Finally, we may conclude, that we have found interesting effects
for the investigated system.  The construction of a
clear connection of these effects to the onset of
 Bose-Einstein condensation as it is known in
statistical physics, is to be made. Further, the problem,
whether our approach is a useful one for the
Bose-Einstein condensation of atomic systems,
{}~\cite{atom}, remains also for future investigations.
 Such studies may be promising since the atoms have to be
trapped with the help of a magnetic trap to cool them below the critical
temperature. The usual theoretical description uses the non-linear Schr\"odinger
equation where a potential is created from $|\psi|^2$. Because of the trap,
plane wave states cannot be utilized. Thus the wave-packet description,
presented in this paper may have a relevance also in the field of
Bose-Einstein condensation of atomic vapors.

\section{Acknowledgments}
Cs. T. would like to thank M. Gyulassy, S. Hegyi, G. Vahtang  and X. N. Wang
for stimulating discussions.
The present study was, in part, supported by the National Science Foundation
(USA) -- Hungarian Academy of Sciences Grant INT 8210278, and by the National
 Scientific Research Fund (OTKA,Hungary) Grant (No.  F4019,
 W01015107 and T024094), by the USA - Hungarian Joint Fund
grant MAKA 378/93 and by an Advanced Research Award from the Fulbright
Foundation. The authors would like to thank
these sponsoring organizations for their aid.
\vfill\eject

\section*{Appendix A}
Here we collect some formulas for another type of factorization of the
$n$-particle density matrix.

For the $n$-particle density matrix we  also may assume, that
it is factorizable in a simple way:
\be
\rho_n(\alpha_1,...,\alpha_n) = \prod_{i=1}^n \, \rho(\alpha_i)
\ee
 This choice of the density matrix is one of
 the possibilities. In the body of the paper we study an other case
where the integrations can be carried out much easier, referred to
as ``model A" in the forthcoming. In this appendix, we explore
the consequences of ``model B", where the integrals are much more
difficult and the calculation can be performed only in some approximate
manner.

Although in this general case one obtains quite complicated integrals
when evaluating the momentum distribution, because of the
overlapping of the wave packets, we can still proceed a little further
with these expressions. Both in case of model A and B, these integrals
for the $m$-pion momentum distribution for fixed multiplicity
$n$ contain the factor given by Eq.~(\ref{e:expec2}).

In case of model B, the two-particle momentum distribution is
given as
\begin{eqnarray}
N^{(n)}_{B,2}({\bk}_1,{\bk}_2)& = &\int \prod_{m=1}^n d\alpha_m  \,
\rho(\alpha_m)
        \times \nonumber \\
\null &\null & \hspace{-2truecm} { \displaystyle \sum_{\sigma^{(n)}}
\sum_{i \ne j = 1}^n
w^*({\bk}_1,\alpha_i) \, w^*({\bk}_2,\alpha_j)
\, w({\bk}_1,\alpha_{\sigma_i}) \, w({\bk}_2,\alpha_{\sigma_j})
\prod_{l=1,l \ne i,j}^n \, \gamma_{l,\sigma_l}
\over \dst
\sum_{\sigma^{(n)}} \prod_{k=1}^n \gamma_{k,\sigma_k} } \label{e:na2c}
\end{eqnarray}
while in case model A, the integrals become somewhat simpler
due to the cancelation of the extra factor in the density matrix with
the denominator of the expectation value in eq. ~\ref{e:expec2}

For model B, the single particle distribution reads as
\bea
N^{(n)}_{B,1}({\bk}_1) & = & \int \prod_{m=1}^n d\alpha_m  \, \rho(\alpha_m)
{ \displaystyle
\sum_{\sigma^{(n)}}
\sum_{i = 1}^n
w^*({\bk}_1,\alpha_i) \,
 w({\bk}_1,\alpha_{\sigma_i}) \,
\prod_{l=1,l \ne i}^n \, \gamma_{l,\sigma_l}
\over \dst
\sum_{\sigma^{(n)}} \prod_{k=1}^n \gamma_{k,\sigma_k} } \label{e:nb1}
\eea

In case of model B, the integrals defining the spectrum and the
two-particle distributions can also be reduced to a
superposition of integrals, which contain only Gaussian factors,
using an expansion of the denominator in the integrand.
 When all $\alpha_i$ -s
are equal, the denominator  reaches its maximum with maximum value
of $n !$. Thus one can expand it in the absolutely convergent series:
\be
{1\over \sum_{\sigma^{(n)}} \prod_{k=1}^n \gamma_{k,\sigma_k} } =
{1\over n! (1-x)} = {1\over n!} \sum_{l=0}^{\infty} x^l
\ee
where we introduced the quantity $ 0 \le x < 1$ with the definition
\be
x = \sum_{\sigma^{(n)}}
{1\over n!}\, \lef 1 - \prod_{k=1}^n \gamma_{k,\sigma_k} \ri
\ee
This expansion is absolutely convergent in terms of $x$ in its domain.
Since $x^n$ contains linear combinations of different powers
of $\gamma$ factors, which are themselves Gaussian factors,
their powers are also Gaussian and since all the remaining factors in
the integrals were Gaussian, we are left with Gaussian integrals.
However, these results are rather complicated even in the
$n=2$ case and we do not include them in the present work.
Due to the greater analytical simplicity of model A,
we have explicitly evaluated integrals only for this latter case,
as was given in the body of the paper.

\vfill\eject

\vfill\eject
\begin{center}
{\bf Figure Captions}
\end{center}

{\bf Fig. 1.}
Solid line stands for the critical
pion multiplicity distribution, $n_c$ as a function of the wave packet
 sizes $\sigma_x$, as indicated by solid line.
The other parameters have the following values:
$ T=120$ MeV, $ R=11$ fm, $ n_0=600. $ 
\medskip

{\bf Fig. 2.}
Pion multiplicity distribution for  wave packet
 sizes $\sigma_x= 1.0, 2.0$ and $ 2.5 $ fm, as indicated by solid,
dashed and dotted line, respectively.
The other parameters have the following values:
$ T=120$  MeV, $R=11 $ fm, $n_0=600. $
\medskip

{\bf Fig. 3.}
Pion multiplicity distribution for  wave packet
 sizes $\sigma_x = 4.0$, $7.0$ and $ 10.0$ fm, as indicated
by solid, dashed and dotted lines, respectively.
The other parameters have the following values:
$ T=120$  MeV, $R=11 $ fm, $n_0=600. $
At a critical size of the wave-packets, the overlap between the packets
will be sufficiently large to start an ``alavanche" of induced emissions,
characterized by a $p_n$ distribution which increases with increasing
values of $n$. (Such distributions can be normalized if they are truncated
at large values of $n$).
As the spatial width of the wave-packets is increased further,
they become more and more similar to plane-waves in momentum space and
the overlap of the wave-functions is decreased. The multiplicity
distribution is shifted back towards the Poissonian limit.
\medskip

{\bf Fig. 4.}
Exclusive single particle momentum distribution,
$P_1^{(600)}(\bk)$ for  a sub-set of events with a fixed multiplicity
of $n = 600$.
 Eq.~(\ref{e:d.p.x}) is plotted for  wave packet
 sizes $\sigma_x = 1.0, 2.0$ and $ 2.5$  fm, as indicated by solid,
 dashed and dotted lines, respectively.
The other parameters have the following values:
$ T=120$  MeV, $R=11 $ fm, $n_0=600. $
The x-axis is scaled with $ \bk^2 $.
Note that the effective slope parameter changes with the variation
of the wave-packet size. As the critical overlap is approached, the
low momentum peak becomes more and more pronounced in the spectrum.
\medskip

{\bf Fig. 5.}
Exclusive single particle momentum distribution,
$P_1^{(600)}(\bk)$ as given by
 eq.(\ref{e:d.p.x}) for  wave packet
 sizes $\sigma= 4.0$, $7.0$ and $ 10.0$ fm, as indicated
by solid, dashed and dotted lines, respectively.
The other parameters have the following values:
$ T=120$ MeV, $R=11$ fm,$ n_0=600.$ The x-axis is scaled with $ \bk^2 $.
 The slope parameter at high momentum is hardly changed when
$\sigma_x$ is varied in this range, but the low momentum enhancement
is strong only if the overlap of the wave-packets approaches the critical
value.
\medskip

{\bf Fig. 6.}
Two-particle exclusive correlation
 function,  as given by eq.(\ref{e:d.x.c}),
	for wave packet
 sizes $ \sigma_x = 1.0$, $2.5$ and $ 10.0$ fm is plotted with
 solid, dashed and dotted lines,
respectively. The mean momentum is fixed to
$ | \bk_1 + \bk_2 | / 2 = 100$  MeV.  The relative momentum
 is parallel to the mean momentum, corresponding to the {\it out}
 direction for spherically symmetric systems.
 The x-axes is scaled with the relative momentum, $ | \bk_1 - \bk_2 | =
	\Delta\bk_{out}. $
 The number of pions was fixed to $n=700$. The ``virtual radii"
are $R_v(\sigma_x=1.0$ fm$ ) = 13.2 $ fm,
   $R_v(\sigma_x=2.0$  fm $)= 11.6$ fm, $R_v(\sigma_x=2.5$ fm $)= 15.2 $ fm.
\medskip

{\bf Fig. 7.}
 Two-particle exclusive  correlation function,
 $C^{(800)}_2(\bk_1,\bk_2)$ is plotted, as given by eq.(\ref{e:d.x.c}),
 for wave packet sizes $ \sigma_x = 1.0$, $2.5$, $4.0 $  and $ 15.0 $ fm
 with solid, dashed, dotted and dense-dotted lines, respectively.
 The mean momentum is fixed to $ | \bk_1 + \bk_2 | / 2 = 50$ MeV.
 The two momenta are parallel (out component).
 The x-axis is scaled with the momentum difference, $ | \bk_1 - \bk_2 | . $
 The actual number of pions was set to $n=800$. The virtual radii
 are $R_v$($\sigma_x=1.0$ fm ) = 12.33 fm,
 $R_v$($\sigma_x=2.5$ fm ) = 8.57 fm,
 $R_v$($\sigma_x=4.0$ fm ) = 8.22 fm, $R_v$($\sigma_x = 15.0$ fm ) = 17.4 fm.
\medskip

\vfill\eject
\vfill\eject

\vfill\eject

\vfill\eject

\vfill\eject

\vfill\eject

\vfill\eject

\end{document}